\pdfoutput=1
\documentclass{aa}  
\usepackage{graphicx}
\usepackage{txfonts}
\usepackage[breaklinks=true]{hyperref}
\usepackage{natbib,twoopt}
\usepackage[breaklinks=true]{hyperref} %% to avoid \citeads line fills
\bibpunct{(}{)}{;}{a}{}{,}             %% natbib format for A&A and ApJ
\usepackage{multirow}
\usepackage{upgreek}
\usepackage{stfloats}

\newcommand{\wtb}{WASP-12\,b}

\newcommand{\hdt}{HD\,209458}
\newcommand{\hdtb}{HD\,209458\,b}
\newcommand{\hdo}{HD\,189733}
\newcommand{\hdob}{HD\,189733\,b}
\newcommand{\gjfb}{GJ\,436\,b}

\newcommand{\ergs}{erg\,s$^{-1}$}

\newcommand{\kms}{km\,s$^{-1}$}
\newcommand{\ms}{m\,s$^{-1}$}

\newcommand{\Mjup}{M$_{\rm jup}$}
\newcommand{\Rpl}{R$_{\rm p}$}

\newcommand{\lya}{Ly$\alpha$}

\newcommand{\lmain}{\ion{He}{i} $\uplambda10830$~\AA{}}
\newcommand{\lminor}{\ion{He}{i} $\uplambda10829$~\AA{}}
\newcommand{\lfive}{\ion{He}{i} $\uplambda5876$~\AA{}}
\newcommand{\hei}[0]{\ion{He}{i}}

\begin{document} 
   
   \title{Detection of \lmain{} absorption on HD~189733\,b\\
          with CARMENES high-resolution transmission spectroscopy}

   \titlerunning{Detection of \lmain{} absorption on HD~189733\,b}
   
   \author{M.~Salz\inst{1}
    \and S.~Czesla\inst{1} 
    \and P.~C.~Schneider\inst{1}
    \and E.~Nagel\inst{1}
    \and J.~H.~M.~M.~Schmitt\inst{1}
    \and L.~Nortmann\inst{2,3}
    \and F.~J.~Alonso-Floriano\inst{4}
    \and M.~L\'opez-Puertas\inst{5}
    \and M.~Lamp\'on\inst{5}
    \and F.~F.~Bauer\inst{5,6}
    \and I.~A.~G.~Snellen\inst{4}
    \and E.~Pall\'e\inst{2,3}
    \and J.~A.~Caballero\inst{7}
    \and F.~Yan\inst{8}
    \and G.~Chen\inst{2,3,9}
    \and J.~Sanz-Forcada\inst{7}
    \and P.~J.~Amado\inst{5}
    \and A.~Quirrenbach\inst{10}
    \and I.~Ribas\inst{11,12}
    \and A.~Reiners\inst{6}
    \and V.~J.~S.~B\'ejar\inst{2,3}
    \and N.~Casasayas-Barris\inst{2,3}
    \and M.~Cort\'es-Contreras\inst{7}
    \and S.~Dreizler\inst{6}
    \and E.~W.~Guenther\inst{13}
    \and T.~Henning\inst{8}
    \and S.~V.~Jeffers\inst{6}
    \and A.~Kaminski\inst{10}
    \and M.~K\"urster\inst{8}
    \and M.~Lafarga\inst{11}
    \and L.~M.~Lara\inst{5}
    \and K.~Molaverdikhani\inst{8}
    \and D.~Montes\inst{14}
    \and J.~C.~Morales\inst{11}
    \and A.~S\'anchez-L\'opez\inst{5}
    \and W.~Seifert\inst{10}
    \and M.~R.~Zapatero~Osorio\inst{15}
    \and M.~Zechmeister\inst{6}
}

   \authorrunning{Salz et al.}
\institute{
% 1
Hamburger Sternwarte, Universit\"at Hamburg, Gojenbergsweg 112, 21029 Hamburg, Germany
\and
% 2
Instituto de Astrof\'isica de Canarias, V\'ia L\'actea s/n, 38205 La Laguna, Tenerife, Spain.
\and
% 3
Departamento de Astrof\'isica, Universidad de La Laguna, 38206 La Laguna, Tenerife, Spain.
\and
% 4
Leiden Observatory, Leiden University, Postbus 9513, 2300 RA, Leiden, The Netherlands
\and
% 5
Instituto de Astrof\'isica de Andaluc\'ia (IAA-CSIC), Glorieta de la Astronom\'ia s/n, 18008 Granada, Spain
\and
% 6
Institut f\"ur Astrophysik, Georg-August-Universit\"at, Friedrich-Hund-Platz 1, 37077 G\"ottingen, Germany
\and
% 7
Centro de Astrobiolog\'ia, CSIC-INTA, ESAC campus, Camino bajo del castillo s/n, 28692 Villanueva de la Ca\~nada, Madrid, Spain
\and
% 8
Max-Planck-Institut f\"ur Astronomie, K\"onigstuhl 17, 69117 Heidelberg, Germany
\and
% 9
Key Laboratory of Planetary Sciences, Purple Mountain Observatory, Chinese Academy of Sciences, Nanjing 210008, China
\and
% 10
Landessternwarte, Zentrum f\"ur Astronomie der Universit\"at Heidelberg, K\"onigstuhl 12, 69117 Heidelberg, Germany
\and
% 11
Institut de Ci\`encies de l'Espai (ICE, CSIC), Campus UAB, C/ de Can Magrans s/n, E-08193 Cerdanyola del Vall\`es, Spain 
\and
% 12
Institut d'Estudis Espacials de Catalunya (IEEC), 08034 Barcelona, Spain
\and
% 13
Th\"uringer Landessternwarte Tautenburg, Sternwarte 5, 07778 Tautenburg, Germany
\and
% 14
Departamento de Astrof\'isica y Ciencias de la Atm\'osfera, Facultad de Ciencias F\'isicas, Universidad Complutense de Madrid, 28040 Madrid, Spain
\and
% 15
Centro de Astrobiolog\'ia, CSIC-INTA, Carretera de Ajalvir km 4, E-28850 Torrej\'on de Ardoz, Madrid, Spain
}

   \date{Received 21 June 2018; accepted 2 November 2018}

% \abstract{}{}{}{}{} 
% 5 {} token are mandatory
% Context, Aims, Methods, Results, Conclusions

   \abstract{
We present three transit observations of \hdob{} obtained with the high-resolution spectrograph CARMENES at Calar Alto.
A strong absorption signal is detected in the near-infrared \hei{} triplet at 10830~\AA{} in all three transits.
During mid-transit, the mean absorption level is $0.88\pm0.04$\,\%
measured in a $\pm$10~\kms{} range at a net blueshift of $-3.5\pm0.4$~\kms{} (10829.84--10830.57~\AA{}). The absorption signal exhibits radial velocities of $+6.5\pm3.1$~\kms{} and $-12.6\pm1.0$~\kms\ during ingress and egress, respectively; all radial velocities are measured in the planetary rest frame. We show that stellar activity related pseudo-signals interfere with the planetary atmospheric absorption signal. They could contribute as much as 80\% of the observed signal and might also affect the observed radial velocity signature, but pseudo-signals are very unlikely to explain the entire signal. 
The observed line ratio between the two unresolved and the third line of the \hei{} triplet is $2.8\pm0.2$, which strongly deviates from the value expected for an optically thin atmospheres.
When interpreted in terms of absorption in the planetary atmosphere, this favors a compact helium atmosphere with an extent of only 0.2 planetary radii and a substantial column density on the order of $4\times 10^{12}$~cm$^{-2}$.
The observed radial velocities can be understood either in terms of atmospheric circulation with equatorial superrotation or as a sign of an asymmetric atmospheric component of evaporating material. 
We detect no clear signature of ongoing evaporation, like pre- or post-transit absorption, which could indicate material beyond the planetary Roche lobe, or radial velocities in excess of the escape velocity. These findings do not contradict planetary evaporation, but only show that the detected helium absorption in \hdob{} does not trace the atmospheric layers that show pronounced escape signatures.
     }


   \maketitle

%_______________________________________________________________________________
%_______________________________________________________________________________
\section{Introduction}\label{SectIntroduction}
%_______________________________________________________________________________
%_______________________________________________________________________________
The atmospheres of close-in planets are exposed to intense high-energy irradiation by their host stars. Stellar extreme-UV and X-ray photons deposit large amounts of energy high up in the planetary atmosphere, capable of powering planetary evaporation with supersonic wind speeds of around 10~\kms{} \citep[e.g.,][]{Lammer2003, Watson1981, Salz2016}. 
Such radiation-induced planetary mass loss may be strong enough to completely evaporate the gaseous envelopes of small planets \citep{Lecavelier2004}, which would explain the detected population of hot super-Earths \citep{Lundkvist2016, Fulton2017}.

An extended hydrogen atmosphere around KELT-9\,b was recently detected via optical H$\alpha$ transit spectroscopy with CARMENES \citep{Yan2018}. To date, the strongest observational evidence for the existence of planetary evaporation winds comes from \lya{} and UV transit spectroscopy. Prominent examples are the planets \hdtb{}, HD\,189733\,b, \wtb{}, and GJ\,436\,b
\citep{Vidal2003, Vidal2004, Vidal2008, Vidal2013, Ehrenreich2008, BenJaffel2010, Linsky2010, Ballester2015,
Lecavelier2010, Lecavelier2012, Bourrier2013, BenJaffel2013,
Fossati2010, Haswell2012, Kulow2014, Ehrenreich2015, Lavie2017}.
\lya{} observations can only be obtained using space-borne instrumentation and, more importantly, interstellar material absorbs the \lya{} line core even for the nearest stars. This absorption suppresses any signal at velocities of around 10~\kms{}, which is the characteristic speed of supersonic evaporation close to the planet.
Therefore, alternative diagnostics for planetary winds are highly desirable.

\cite{Seager2000} were the first to emphasize the potential of the \lmain{} triplet lines to study upper atmospheric layers, which are the launching region of the planetary wind. The triplet is composed of two closely spaced lines with central wavelengths\footnote{We use air wavelengths throughout the manuscript.} of 10830.33 and 10830.25~\AA{} and a third weaker line at 10829.09~\AA. These lines are accessible from the ground and are not affected by interstellar absorption. For example, along one line of sight, \citet{Indriolo2009} derived an upper limit of $3.2\times 10^9$~cm$^{-2}$ for the interstellar \hei{} triplet state column density, which is two to three orders of magnitude lower than the column densities derived in the following. 
Recently, \citet{Oklopcic2018} used 1D models of the escaping atmospheres of \gjfb{} and \hdtb{}, including all sources and sinks of the triplet ground state, and showed that planetary absorption in the \lmain{} lines could reach several percent.

Absorption in the \hei{} triplet lines in the stellar atmosphere is related to stellar activity features \citep[e.g.,][]{Zarro1986, Sanz2008}, which can cause serious complications for exoplanet transit observations.
In the solar context, the lines have been studied in detail by \cite{Avrett1994} and  \cite{Mauas2005}. The Sun shows a highly inhomogeneous surface distribution of \lmain{} absorption \citep[see Fig.~1 of][]{Andretta2017}, with absorption practically restricted to active regions \citep{Avrett1994}. Accordingly, the disk-integrated absorption reveals 10\% rotational variation and 30\% variation over activity cycles \citep{Harvey1994}.

While a search  with the VLT/ISAAC instrument for atmospheric \lmain{} absorption of the hot Jupiter orbiting the inactive host star \hdt{} resulted in an upper limit of 0.5\% in a 3~\AA{} wide window \citep{Moutou2003}, the signal has now been detected in WASP-107\,b  with Wide Field Camera 3 on board the Hubble Space Telescope. Here, 0.05\% excess absorption was observed over a 98~\AA{} wide window \citep{Spake2018}. In \citet{Lisa2018}, we report $3.6\pm0.2$\,\% \hei{} absorption observed during the transit of WASP-69\,b at high spectral resolution with the CARMENES spectrograph. Here, we present the detection and analysis of \hei{} absorption in the \hdo{} system.

%_______________________________________________________________________________
%_______________________________________________________________________________
\section{The \hdo{} exoplanetary system}\label{SectTarget}
%_______________________________________________________________________________
%_______________________________________________________________________________

\begin{table}[h]
  \caption{Adopted system parameters of \hdo{}}
  \label{TabPara}
  \centering
  \begin{tabular}{l c l}
    \hline\hline\vspace{-8pt}\\
     Parameter      & Value & Reference   \\
    \vspace{-10pt}\\\hline\vspace{-10pt}\\
     $\alpha$ [J2015.5]            & 20:00:43.70                       &  {\it Gaia} DR2\tablefootmark{a} \\\vspace{-10pt}\\
     $\delta$ [J2015.5]            & +22:42:35.3                       &  {\it Gaia} DR2\tablefootmark{a} \\\vspace{-10pt}\\
     $d$                          & $19.775\pm 0.013$~pc              &  {\it Gaia} DR2\tablefootmark{a} \\\vspace{-10pt}\\
     $R_{\rm S}$\tablefootmark{b} & $0.805\,(16)~R_\sun$              &  \citet{Boyajian2015} \\\vspace{-10pt}\\
     $M_{\rm S}$                  & $0.846^{+0.06}_{-0.049}~M_\sun$   &  \citet{deKok2013} \\\vspace{-10pt}\\
     $P_{\rm rot}$                & $11.953\,(9)$~d                   &  \citet{Henry2008} \\\vspace{-10pt}\\     
     $K_{\rm S}$                  & $201.96^{+1.07}_{-0.63}$~m\,s$^{-1}$ &  \citet{Triaud2009} \\\vspace{-10pt}\\
     $v_{\rm sys}$                & $-2.361$\,(3)~\kms{}              &  \citet{Bouchy2005} \\\vspace{-10pt}\\
     $T_0$ [BJD$_{\rm TDB}$]      & \hspace{-1mm}2453955.5255511\,(88)\hspace{-1mm} & \citet{Baluev2015} \\\vspace{-10pt}\\
     $P_{\rm orb}$                & 2.218575200\,(77)~d               & \citet{Baluev2015} \\\vspace{-10pt}\\
     $b$\tablefootmark{c}         & 0.6636\,(19)                      & \citet{Baluev2015} \\\vspace{-10pt}\\
     $R_{\rm p}/R_{\rm s}$        & 0.15712\,(40)                     & \citet{Baluev2015} \\\vspace{-10pt}\\
     $M_{\rm p}$                  & $1.162^{+0.058}_{-0.039}$~\Mjup{} &  \citet{deKok2013} \\\vspace{-10pt}\\
     $a/R_{\rm s}$                & 8.863\,(20)                       & \citet{Agol2010} \\\vspace{-10pt}\\
     $i$                          & 85.710\,(24)\degr{}               & \citet{Agol2010} \\\vspace{-10pt}\\
     $K_{\rm p}$                  & 162.2\,(3.3)~\kms{}               &  this work\tablefootmark{d} \\
    \vspace{-9pt}\\\hline
  \end{tabular}
  \tablefoot{
             \tablefoottext{a}{\citet[][]{Gaia2016A, Gaia2018}}
             \tablefoottext{b}{Measured by interferometry. The stellar radius determines the absolute dimension of the semimajor axis and the
             planetary radial velocity half amplitude $K_{\rm p}$.}
             \tablefoottext{c}{Impact parameter.}
             \tablefoottext{d}{$K_{\rm p}=2\pi a/P_{\rm orb}\,\sin i$.}
            }
\end{table}

The hot Jupiter \hdob{} is among the best-studied planets to date.  The 1.16~\Mjup{} mass planet orbits an active K dwarf with a period of 2.2~days \citep{Bouchy2005}; see Table~\ref{TabPara} for details. Its atmospheric transmission is likely dominated by Rayleigh scattering in the wavelength range from 3000 to 10\,000~\AA{} \citep{Pont2008, Pont2013, Lecavelier2008, Sing2009, Sing2011, Sing2016, Gibson2012}. However, the contribution of unocculted stellar spots to the alleged Rayleigh scattering slope remains uncertain \citep{McCullough2014}.
In lower atmospheric layers, absorption of carbon monoxide \citep{deKok2013, Rodler2013, Brogi2016} and water \citep{Birkby2013, McCullough2014, Brogi2018} has been detected. Measurements of sodium absorption were used to reconstruct the atmospheric temperature-pressure profile up to the lower thermosphere \citep{Redfield2008, Huitson2012, Wyttenbach2015, Louden2015, Sara2017}. A reported detection of planetary H$\alpha$ absorption by \citet{Jensen2012} remains difficult to interpret due to the confounding spectral effects of stellar variability \citep{Barnes2016, Cauley2017}.

\hdo{} is an active star with strong emission cores in the \ion{Ca}{ii} H\&K lines, resulting in a high value of $0.508$ for the Mount Wilson S-index \citep{Baliunas1995, Knutson2010}. The star shows frequent flaring at optical and X-ray wavelengths \citep{Pilliteri2014, Klocova2017} and an overall X-ray luminosity of $\approx$~2~$\times$ 10$^{28}$~\mbox{erg\,s$^{-1}$} \citep{Huensch1999}, which 
places it in the top decile of the X-ray luminosity distribution function \citep{schmitt1995}. \citet{Sanz2011} reconstructed an extreme UV luminosity of $3\times10^{28}$~\ergs{}, which implies substantial high-energy irradiation levels on \hdob{} that ought to trigger an evaporative wind in the upper planetary atmosphere \citep{Salz2016}. The escape of this upper atmosphere has been detected through hydrogen \lya{} and oxygen absorption \citep{Lecavelier2010, Lecavelier2012, Bourrier2013, BenJaffel2013}, and \citet{Poppenhaeger2013} proposed a tentative $6-8$\% deep X-ray transit. These findings make \hdob{} a promising candidate to search for \lmain{} absorption.

%_______________________________________________________________________________
%_______________________________________________________________________________
\section{Observations and data analysis}\label{SectCarmenes}
%_______________________________________________________________________________
%_______________________________________________________________________________
We analyzed three spectral transit time series\footnote{
Reduced spectra are available at the Calar Alto Archive (\url{http://caha.sdc.cab.inta-csic.es/calto/}).} of \hdo{} taken on 8 Aug 2016, 17 Sept 2016, and 7 Sept 2017, in the following referred to as night 1, 2, and 3. The observations were taken with the CARMENES spectrograph, mounted at the 3.5~m telescope at the Calar Alto Observatory; see  \cite{Quirrenbach2016} for a detailed description. The CARMENES spectrographs simultaneously cover the visual and near-infrared (NIR) ranges ($5500 - 9600$~\AA{} and $9600 - 17\,200$~\AA) with a nominal resolution of $94\,600$ and $80\,400$, respectively, and a sampling of 2.3~pixel per resolution element around the \hei{} triplet.
The two independent channels are housed in vacuum tanks to optimize radial velocity precision. The NIR spectrograph is fed by two fibers. In our configuration, fiber~A carried the light from the target and fiber~B was used to obtain simultaneous sky spectra. The data reduction was carried out with the pipeline \citep[CARACAL v2.10,][]{Caballero2016}.

\begin{table*}
\centering
\caption{Observing details \label{tab:obs_details}}
  \begin{tabular}{lcccccccc}
    \hline
    \hline
    Night\hspace{-3mm} & Date & Proposal ID & Principal & Calar Alto & Nr. of & S/N\tablefootmark{b} at  & Pre-/post-  & Mid-transit time\\
          &      & & investigator & Archive ID & spectra\tablefootmark{a} & 10830\,\AA{} & transit (h) & (BJD TDB) \\ 
    \hline
    1 & 2016-08-08 & H16-3.5-024 & P. J. Amado     & 246903-246952 & 45 (1/1) & 160 & 1.0 / 1.0  &  2457609.51891\\
    2 & 2016-09-17 & H16-3.5-024 & P. J. Amado     & 249745-249798 & 50 (4/2) & 210 & 2.3 / 1.2  &  2457649.45326\\
    3 & 2017-09-07 & H16-3.5-022 & J. A. Caballero & 263103-263156 & 46 (1/1) & 240 & 1.7 / 1.0  &  2458004.42529\\
    \hline
  \end{tabular}
  \tablefoot{
             \tablefoottext{a}{Number of spectra neglected at the beginning and end of the night in parentheses.}
             \tablefoottext{b}{One divided by the nightly average standard deviation of the residual spectra in the continuum surrounding the \lmain{} lines.}
            }
\end{table*}

\begin{figure}[t]
  \centering
  \includegraphics[width=\hsize, trim = 2.5mm 2.5mm 3mm 2mm, clip]{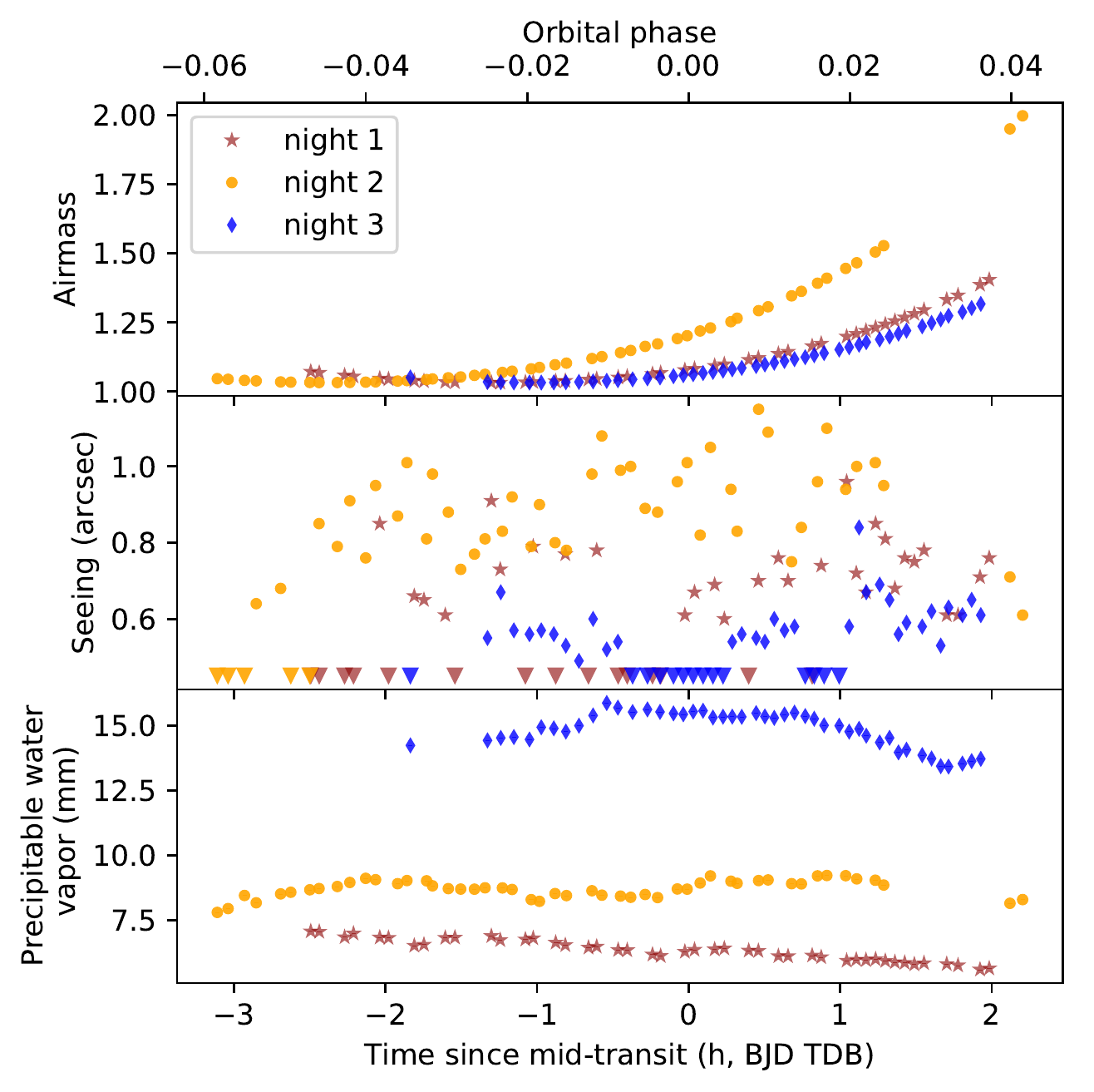}
  \caption{Airmass, seeing, and column of precipitable water vapor during 
           the observations.
           Unresolved seeing is indicated by triangles. 
           The error bars for the water column, which was derived in the telluric
           correction, are smaller than the marker size.
           }
  \label{fig:obscond}
\end{figure}

The exposure time of our spectra was 198\,s throughout the campaign. Further details on the observations and the observing conditions are provided in Table~\ref{tab:obs_details} and Fig.~\ref{fig:obscond}. The airmass was mostly below 1.5 and the typical seeing was better than 1\arcsec{}. During night~3 the column of water vapor was higher, but this night offered the best seeing conditions and resulted in the highest signal-to-noise ratio (S/N) per spectrum. The S/N during the first two nights may have been affected by stability issues with the NIR channel, which is evident in the NIR radial velocity measurements (see Appendix~\ref{Sect:RM}).

Scrutinizing the spectral time series, we identified a total of four spectral regions that exhibited excessive spectral variations in all three nights. These spectral anomalies were likely caused by bad detector pixels and were, therefore, discarded in our analysis. Fortunately, none of the affected regions overlapped with the \hei{} lines (shaded regions in Fig.~\ref{fig:tsp}).

\subsection{Telluric correction}

Although the line cores of the \lmain{} lines are not blended with telluric lines, the spectral region is contaminated with water vapor absorption originating from the Earth's atmosphere. We used the \url{molecfit} software in version 1.2.0 \citep{Smette2015, Kausch2015} to remove the telluric contribution from each individual CARMENES spectrum. A nightly mid-latitude ($45^{\circ}$) reference model atmosphere\footnote{\url{http://eodg.atm.ox.ac.uk/RFM/atm/}} and the Global Data Assimilation System (GDAS) profiles for the location of the Calar Alto Observatory were used to create atmospheric temperature-pressure profiles for the three nights, which are required for the line-by-line radiative transfer in \url{molecfit}.  We included O$_2$, CO$_2$, and CH$_4$ in our transmission model with fixed abundances, taken from the reference atmosphere, and fitted the precipitable water vapor.

As reported by \citet{Allart2017}, the choice of the optimization ranges is crucial to derive precise transmission models with \url{molecfit}. We selected nine 50--100\,$\AA$ broad wavelength intervals evenly distributed over the CARMENES NIR channel, taking full advantage of the large amount of telluric lines contained in this spectral region. These intervals exhibit few stellar lines, a well determined continuum level, and comprise various deep but unsaturated telluric absorption lines. Stellar features were identified and masked using a high-resolution synthetic stellar spectrum \citep[PHOENIX,][]{Husser2013} with $T_{\mathrm{eff}}=4700\,$K, $\log g=4.5\,$dex, and solar metallicity. In addition, we excluded wavelength ranges with sky emission features. The instrumental line spread function was determined using hollow cathode lamp spectra.
Based on the best-fit parameters derived by \url{molecfit}, we finally generated a transmission model for the entire wavelength range of CARMENES and corrected each science spectrum. 

\begin{figure}[tb]
  \centering
  \includegraphics[draft=false, width=\hsize, trim = 1.1mm 7.5mm 3mm 0mm, clip]{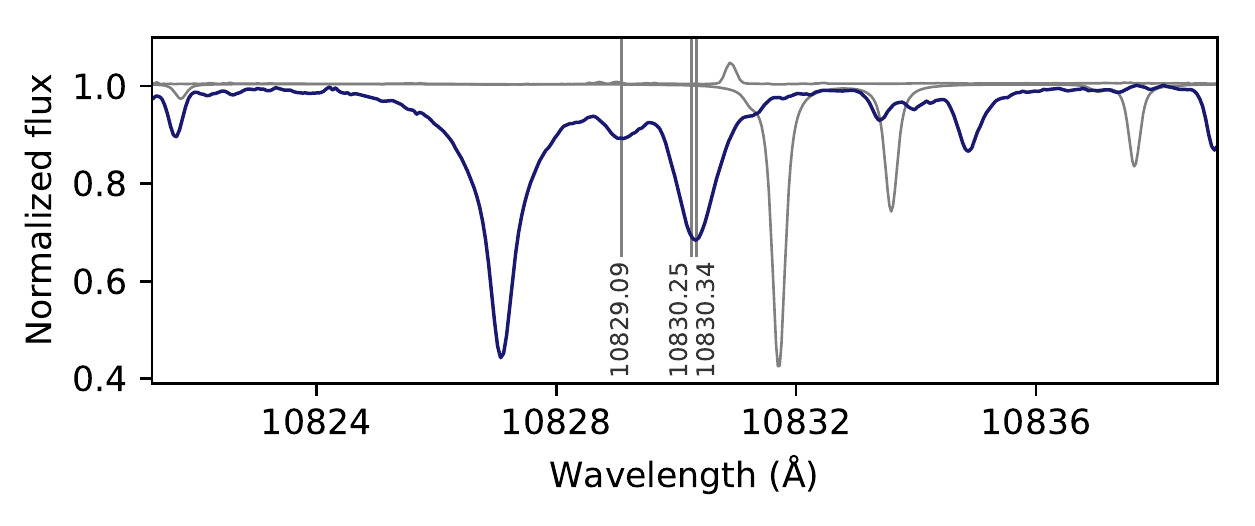}\\
  \includegraphics[draft=false, width=\hsize, trim = 2.5mm 13mm 3mm 3mm, clip]{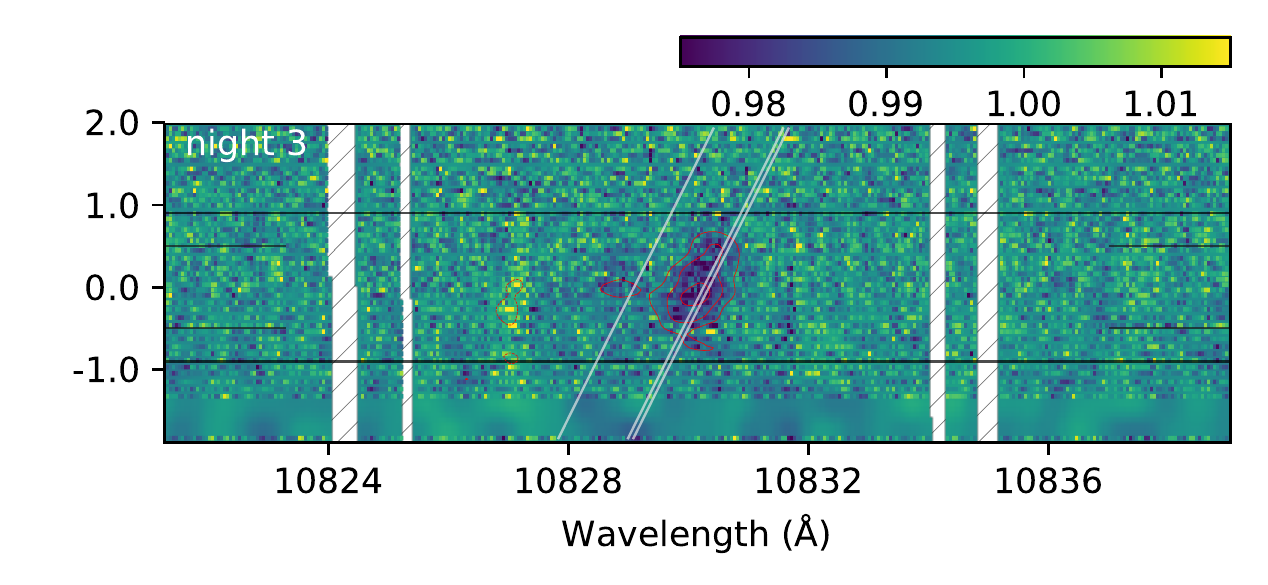}\\
  \includegraphics[draft=false, width=\hsize, trim = 2.5mm 13mm 3mm 14mm, clip]{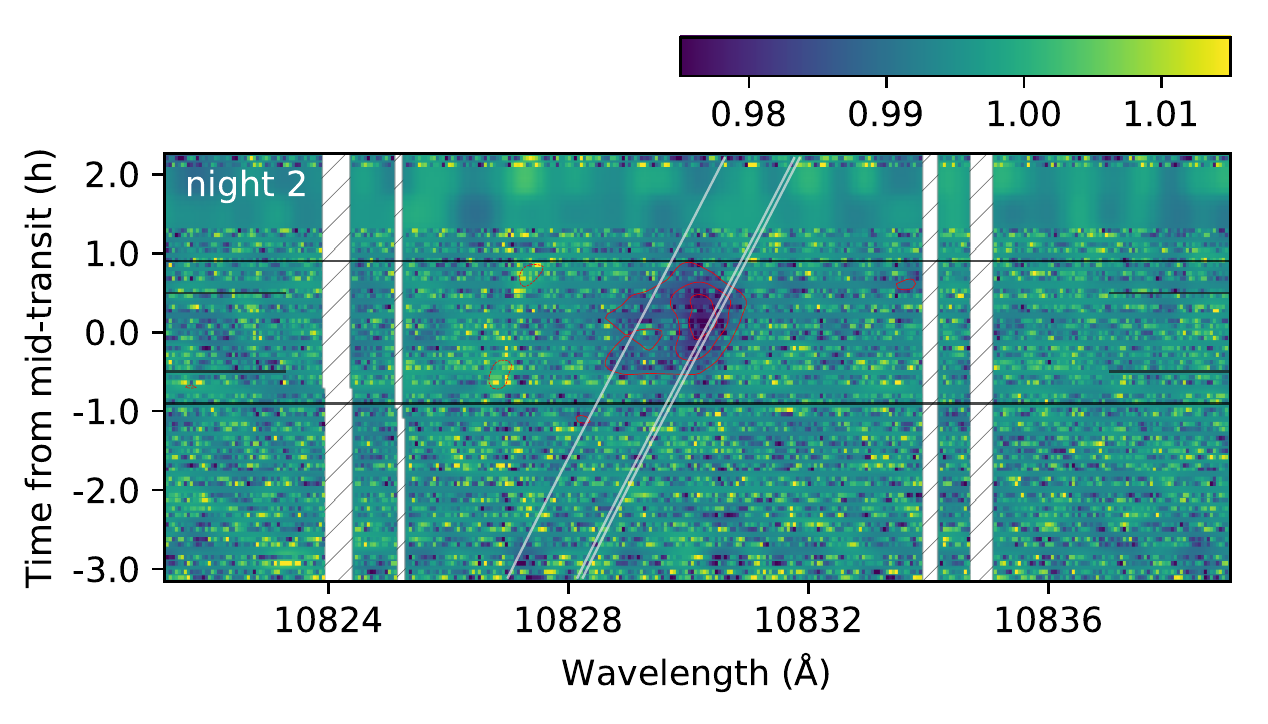}\\
  \includegraphics[draft=false, width=\hsize, trim = 2.5mm 0mm 3mm 12.5mm, clip]{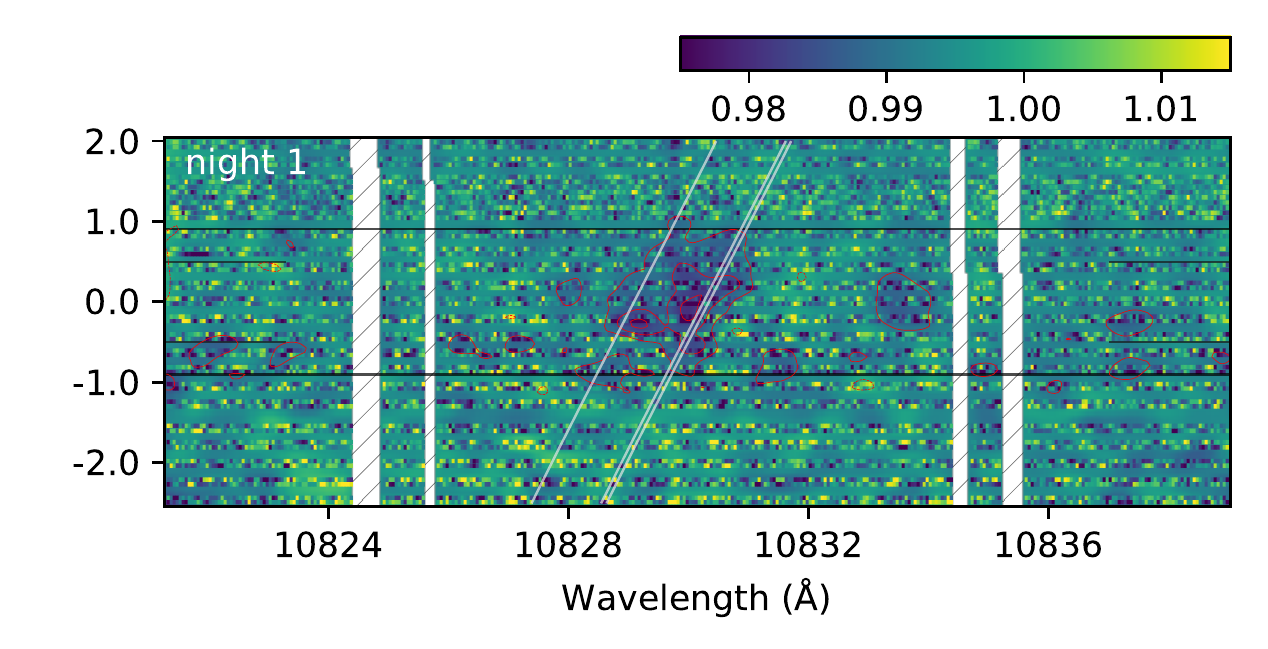}
  \caption{{\it Top panel:} Out-of-transit spectrum around the \lmain{} lines
           after removal of telluric lines and averaged over the three observing nights.
           The triplet line positions are marked by vertical lines labeled by
           their central wavelength.
           A telluric absorption spectrum and the average
           telluric emission spectrum from night~3 are depicted by gray lines.
           {\it Lower panels:} Time series of residual spectra for three
           transits of \hdob{} in the stellar rest frame.
           Rows depict individual residual spectra with color-coded intensity;
           time progresses along the y-axis, and row height and gaps correspond to the exposure times.
           We use a linear interpolation over the gaps and apply a Gaussian smoothing and transparency for distinction.
           Masked regions are likely affected by bad pixels.
           First and fourth contacts of the optical transit are marked by horizontal
           lines, and second and third contacts are indicated by shorter vertical lines
           at the figure edges.
           The slanted lines show the position of the \hei{}~triplet lines
           {in the planetary rest frame}. 
           Red solid and dashed contours are respectively 0.33\% negative and positive variations from mean.
           }
  \label{fig:tsp}
\end{figure}

Telluric emission lines were removed using the sky spectrum from fiber~B. We fitted and subtracted the continuum in the sky spectra and subtracted the remaining emission line spectrum from the science spectrum. The comparison of Figs.~\ref{fig:tsp} and \ref{fig:tsp_ntc} shows that the strongest emission line in the red wing of the \hei{} triplet at 10830.91~\AA{} was successfully removed to within the noise level. At any rate, this line can only affect the egress phase, where its position partially overlaps with the \lmain{} lines. Two further telluric emission lines at 10829.01~\AA{} and 10828.73~\AA{} are located such that they could affect the analysis of the weaker triplet component at 10829~\AA{}. However, these lines are a factor of 11 weaker than the line at 10830.91~\AA{}. As we found no significant differences in our results when correcting or masking them, we consider these lines irrelevant.

\subsection{Continuum normalization and stellar rest frame alignment \label{tab:cont_norm}}

The continuum was normalized with a third-order polynomial and the spectra were shifted into the stellar rest frame by correcting for a systematic radial velocity of $-2.361$~\kms{} \citep{Bouchy2005}, Earth's barycentric velocity, and the stellar orbital motion.
The barycentric velocity correction was computed using the {\tt helcorr}
routine\footnote{Implemented in PyAstronomy \url{https://github.com/sczesla/PyAstronomy} and adapted from the REDUCE package \citep{Piskunov2002}.},
and mid-exposure time stamps were converted into Barycentric Julian Dates (BJD) in Barycentric Dynamic Time (TDB) using the {\tt Astropy} time package \citep{Astropy2013}.

For each night, the spectral alignment was controlled using the seven strongest and isolated stellar lines in the vicinity of the \hei{} triplet (see Fig.~\ref{fig:star_rf}):
\ion{Mg}{i} $\lambda$10811.084~\AA{},
\ion{Fe}{i} $\lambda$10818.276~\AA{},
\ion{Si}{i} $\lambda$10827.091~\AA{},
\ion{Ca}{i} $\lambda$10838.970~\AA{},
\ion{Si}{i} $\lambda$10843.854~\AA{},
\ion{Fe}{i} $\lambda$10849.467~\AA{}, and
\ion{Fe}{i} $\lambda$10863.520~\AA{}.
We found an average redshift of 360~\ms{} with respect to the radial velocity of \citet{Bouchy2005}, which was corrected. During night~2, the instrument showed an apparent drift of $\approx$\,200~\ms{}. This drift was modeled for the out-of-transit phases with a second-order polynomial, which was then used to correct the alignment of all spectra during this night.
Radial velocity drifts or offsets can occur because our observations were not obtained with a setup optimized for radial velocity measurements, i.e., the instrumental radial velocity drift was not monitored through the Fabry Perot in the calibration fiber. In the end, we found all seven stellar lines within $\approx$\,200~\ms{} of their nominal central wavelengths and interpolated all spectra onto a common wavelength grid.

\subsection{Residual spectra}\label{SectAnalysis}

For each night, we combined the out-of-transit spectra to construct a master reference spectrum, discarding the first and last few spectra during each night for technical reasons (see Table~\ref{tab:obs_details}). All spectra were then divided by this master out-of-transit spectrum to obtain a time series of residual spectra (see Fig.~\ref{fig:tsp}). In all three observed transits, a pronounced absorption feature attributable to the \lmain{} lines is detected. These results are independent of our telluric emission correction (see Fig.~\ref{fig:tsp_ntc}).
The highest S/N was reached on night 3, during which the radial velocity of the absorption signals clearly shifts along with the orbital radial velocity of the planet in the mid-transit phase. This indicates that the absorption is indeed associated with the hot Jupiter.
The association is less clear on the first two nights. This is could be caused by the reduced instrumental stability and S/N (see Appendix~\ref{Sect:RM}), but could also be related to stellar pseudo-signals interfering with the planetary \lmain{} absorption signals. This possibility is further investigated in Sect.~\ref{SectSpots}.
The Rossiter--McLaughlin effect (RME) in the \lmain{} lines is also superimposed on the residual spectra, but we calculate its amplitude to be smaller than 0.07\%, which is negligible during the mid-transit phase (see Sect.~\ref{SectSpots}).

The out-of-transit stellar \lmain{} lines do not show detectable variation across the three observing nights or within any of them (see Fig.~\ref{fig:tsp}). The cores of the \ion{Ca}{ii} infrared triplet lines, which are well-known activity indicators, show an activity trend during night~1 along with small activity fluctuations, but no clear signature of flaring (see Appendix~\ref{Sect:APP_IRT}).
Figure~\ref{fig:tsp} exhibits a weaker feature associated with the \ion{Si}{i} line at 10827.1~\AA{} that exhibits a radial velocity of only about 100~\ms{} in the stellar rest frame. This feature can be  nicely described  by center-to-limb variations in combination with the RME in the stellar line \citep{Czesla2015}. There is also a slight activity trend in this line that correlates with the trend seen in the \ion{Ca}{ii} infrared triplet lines.
 
\begin{figure}[tb]
  \centering
  \includegraphics[draft=false, width=\hsize, trim = 2.5mm 2.5mm 3mm 2mm, clip]{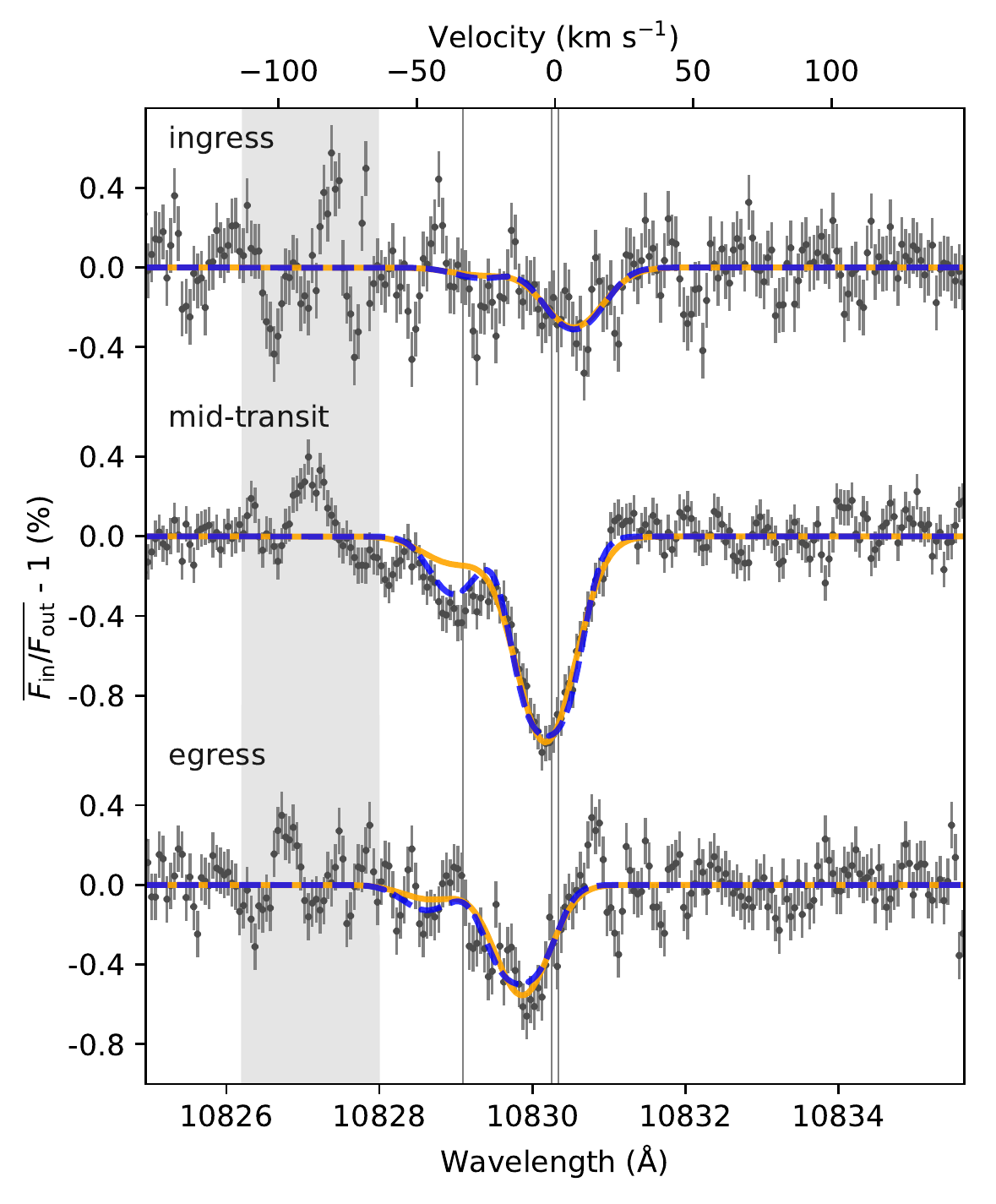}
  \caption{Weighted mean transmission spectrum of the three observed transits split
           into ingress, mid-transit, and egress. 
           The wavelengths of the triplet lines are marked by vertical lines. 
           Compact and extended atmosphere models are depicted for each phase 
           by blue dashed and orange solid lines, respectively (see main text).
           The gray shaded region was masked during the MCMC runs due to 
           residuals in a \ion{Si}{i} line.
           }
  \label{fig:mean}
\end{figure}

Using the ephemeris of \citet{Baluev2015}, we shifted the residual spectra into the planetary rest frame and computed nightly means for the ingress, mid-transit, and egress phases. Observations that start after the first contact and 
have a mid-exposure time before the second contact were included in the ingress phase. An equivalent procedure was used for the egress phase. In total, we have 10 spectra during the ingress, 34 during mid-transit, and 13 during the egress phases.
Because of the different data quality, the three nightly means were combined weighting them by the inverse variance in the surrounding continuum. The obtained mean transmission spectra are displayed in Fig.~\ref{fig:mean}, and the nightly residual spectra and their variation with respect to the mean transmission spectrum are shown in Fig.~\ref{fig:var+lc2}. 

\subsection{Bootstrap analysis of the absorption depth}\label{Sect:APP_absdepth}

\begin{figure}[t]
  \centering
  \includegraphics[width=\hsize, trim =2.5mm 2.5mm 3mm 2mm, clip]{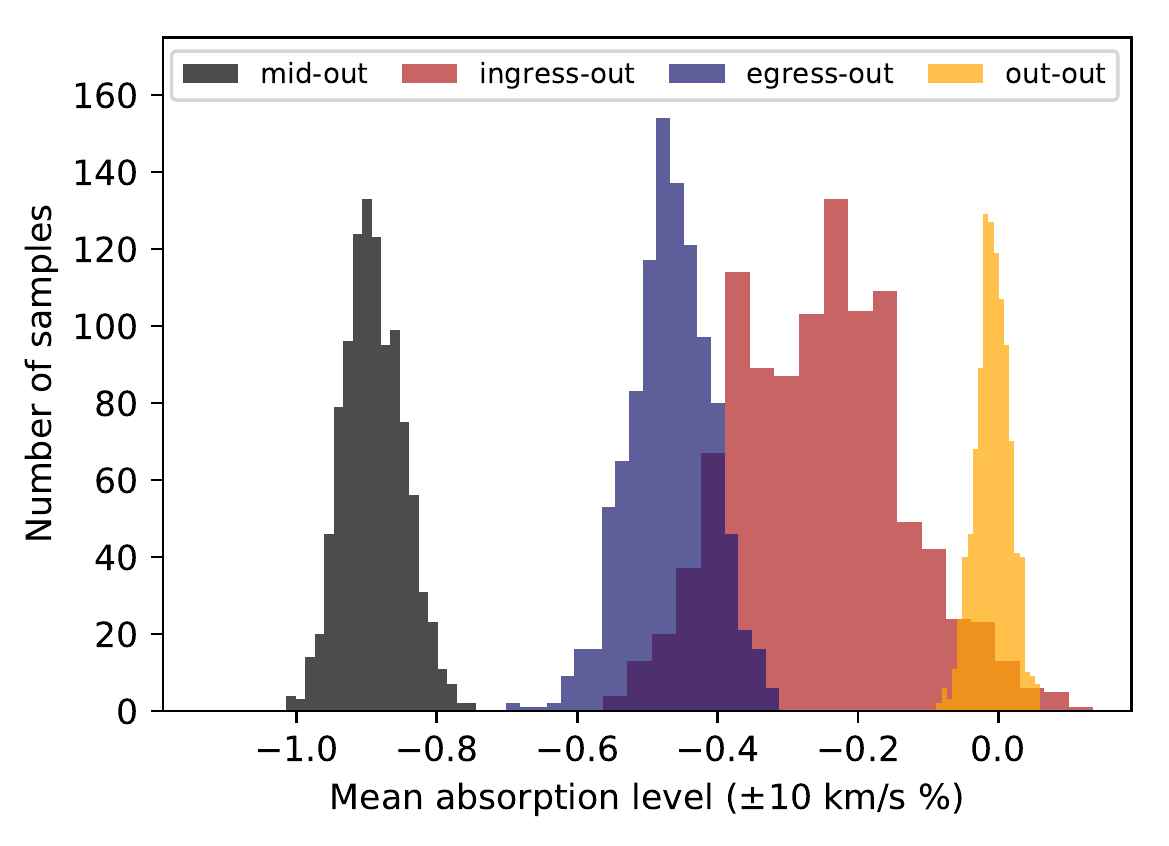}
  \caption{Histograms of the absorption levels in the mean residual
           spectra for the three in-transit
           phases and a control out-of-transit phase during our bootstrap analysis.
           }
  \label{fig:APP_bootstrap}
\end{figure}

For each phase, the mean absorption depth was calculated in a $\pm10$~\kms{} window centered on the shifted signal position determined by our model fitting in Sect.~\ref{SectModel}: $+6.5$~\kms{} for ingress, $-3.5$~\kms{} for mid-transit, and $-12.6$~\kms{} for egress. Errors were determined by a bootstrap method. On average, we randomly drew  half of the spectra from each phase, computed the mean absorption spectrum, and determined the average absorption level. This was performed 1000 times and the resulting histograms are shown in Fig.~\ref{fig:APP_bootstrap}. We find average absorption levels of $0.24\pm0.12$\,\% for the ingress phase, $0.88\pm0.04$\,\% for the mid-transit phase, and $0.46\pm0.06$\,\% for the egress phase. We also randomly drew half of the out-of-transit residual spectra, averaged in the center of the two strong helium lines at 10830~\AA{} in the stellar rest frame, and find zero absorption in this control sample.

The ingress signal is formally only a 2$\sigma$ result and exhibits the largest scatter of the three transit phases. This cannot be caused by residual telluric contamination since the ingress signal does not overlap with any telluric emission or absorption line.

In the line core of the strong \hei{} triplet component, the absorption depth reaches its maximum of $1.04\pm0.09$\%, using the standard deviation of the surrounding continuum as error estimate. If this absorption feature is caused by the planet, its atmosphere must exhibit a radial extent of at least 0.2~\Rpl{}. The absorption signal has a total equivalent width of $12.7\pm0.6$~m\AA{} and  we further derive a ratio of the \lmain{} to $\uplambda10829$~\AA{} triplet components of $2.8\pm0.2$ by fitting Gaussians with free relative strengths. Here the errors are propagated from the bootstrap analysis.

To investigate nightly variability in the absorption depth, we repeated the bootstrap analysis for the mid-transit signals of the individual nights averaging over the full signal ($-70$ to $+30$~\kms{}). We measure absorption levels of $0.41\pm0.04$\,\%, $0.39\pm0.04$\,\%, and $0.35\pm0.04$\,\% for nights 1, 2, and 3 respectively. While the nightly mean transmission spectra suggest some variation at different epochs (see Fig.~\ref{fig:var+lc2}), they do not exceed the rednoise level.

\subsection{Temporal evolution of the \ion{He}{i} signal}
\label{sec:TempEvol}

\begin{figure}[t]
  \centering
  \includegraphics[width=\hsize, trim =2.5mm 2.5mm 3mm 2mm, clip]{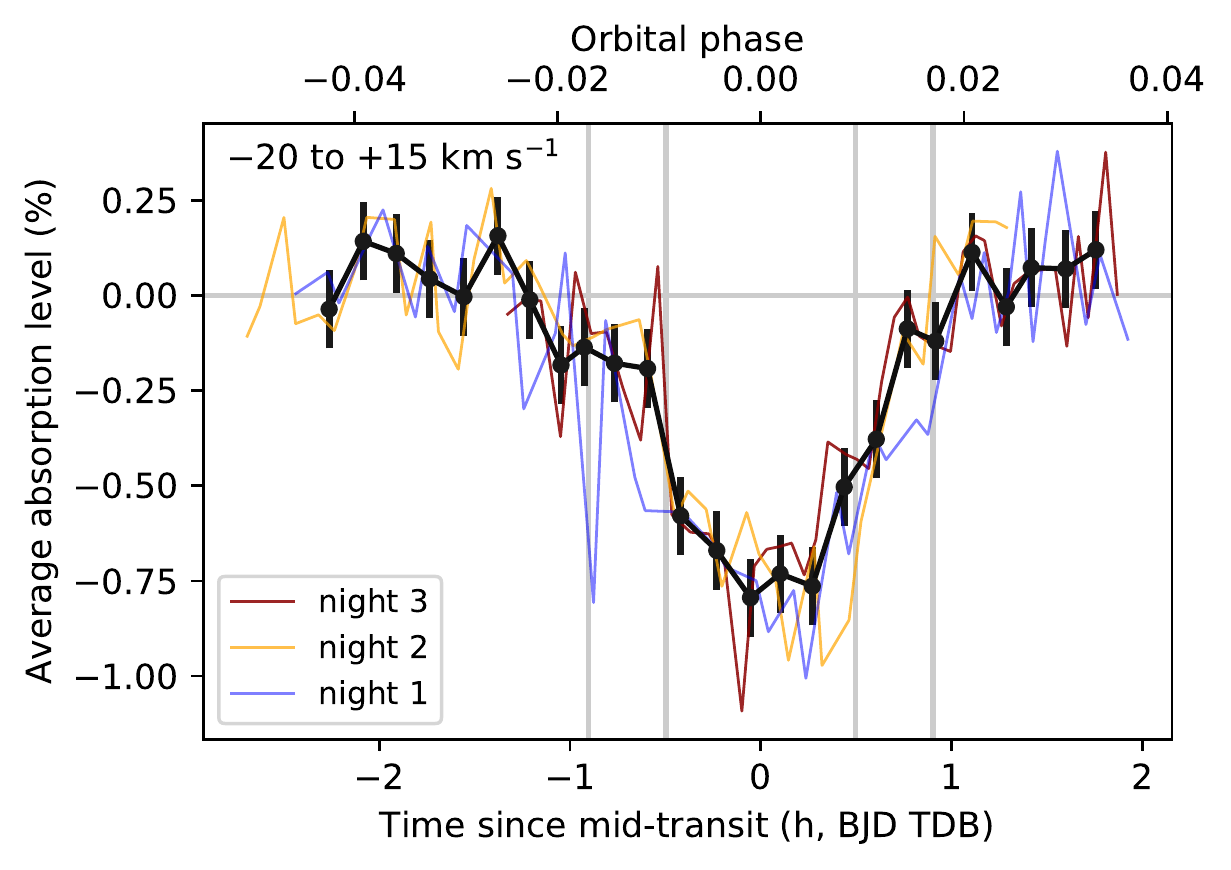}
  \caption{Time evolution of the \lmain{} signal during the three nights.
           Vertical lines mark the contact points of the broad band optical
           transit (T$_1$ to T$_4$).
           The mean transit light curve (black line) is
           shown with error bars.
           }
  \label{fig:lc}
\end{figure}

We computed light curves of the \hei{} signal by averaging over a broad velocity range from $-$20 to $+$15~\kms{} in the planetary rest frame, corresponding to 10829.57 to 10830.84~\AA. This range covers the observed velocity shifts at ingress and egress. The light curves are depicted in  Fig.~\ref{fig:lc}. A light curve centered on the weak triplet component is provided in Fig.~\ref{fig:var+lc2}.
We detect no significant inter-night variations. While some apparent in-transit outliers could be associated with spot crossing events, we consider the evidence immaterial. A mean light curve was constructed by binning the three phased light curves with a temporal resolution of 10~min. The nightly weights were maintained in this procedure. Errors are obtained from the variation in the out-of-transit phase. The resulting time series is nearly symmetric with respect to the transit center. Some pre-transit absorption may be present, but
the evidence remains inconclusive; we find no evidence for post-transit absorption.

%_______________________________________________________________________________
%_______________________________________________________________________________
\section{Discussion}\label{SectDiscussion}
%_______________________________________________________________________________
%_______________________________________________________________________________
Our data show a clear in-transit signal in the \hei{} triplet lines, consistently present in all three spectral transit time series. However, this signal is not necessarily caused only through absorption in a planetary atmosphere because the transit of the opaque planetary disk across an inhomogeneous stellar surface can produce pseudo-absorption and pseudo-emission signals. As a first step toward a better understanding of the presented signal, we study the two extreme cases:  a heterogeneous stellar surface and no planetary \hei{} absorption, and  only planetary absorption.

\subsection{Impact of a spotted stellar surface \label{SectSpots}}

Even without atmosphere, the transit of the planetary disk over stellar surface regions with below-average \hei{} absorption (bright stellar surface patches) produces an apparent absorption signal in the residual spectra (pseudo-absorption). Similarly, pseudo-emission is produced when the planet traverses dark stellar surface regions with strong stellar \hei{} absorption.

The impact of these pseudo-signals can be investigated considering the limiting scenarios. When the planet transits stellar surface patches completely lacking the \hei{} absorption line, the maximum amount of pseudo-absorption is observed. While the amplitude of this signal only depends on the average stellar \lmain{} spectrum and the optical transit depth, it is necessary  to specify the stellar \hei{} spectrum in the eclipsed section of the disk to compute the expected pseudo-emission signal.
To that end, we applied a two-component disk model, consisting only of regions that are either entirely free of absorption or show strong \hei{} absorption.

\subsubsection{Filling factor of dark \hei{} patches}\label{Sect:stellar_he}

\begin{figure}[t]
  \centering
  \includegraphics[width=\hsize, trim = 2.5mm 7.5mm 3mm 2mm, clip]{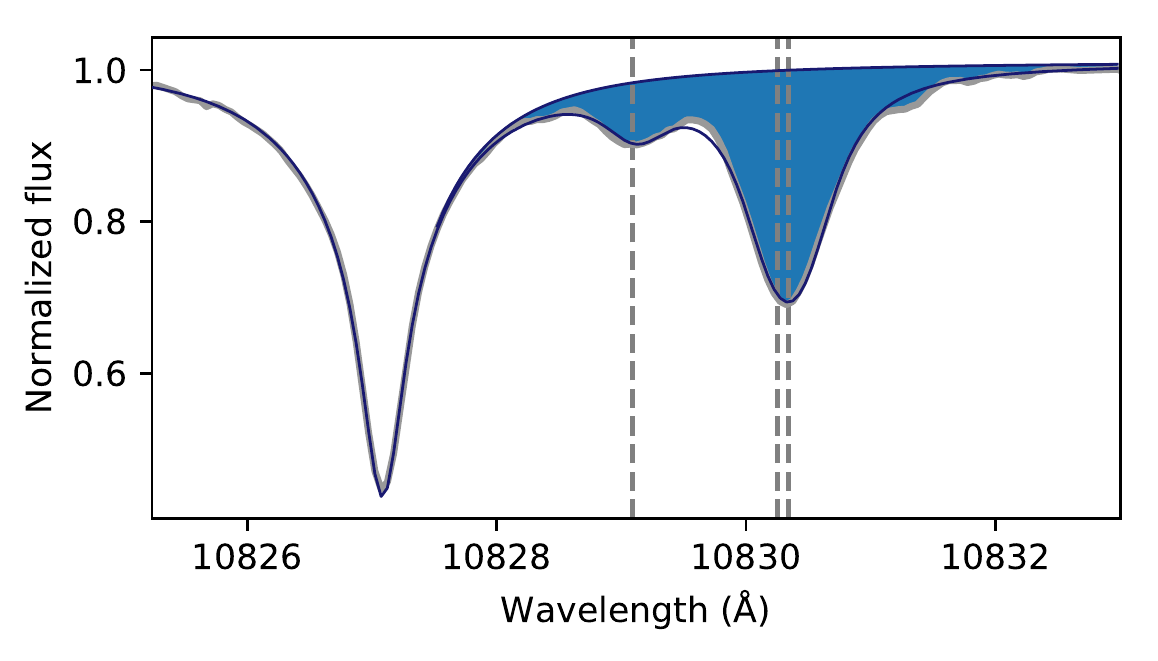}
  \includegraphics[width=\hsize, trim = 2.5mm 2.5mm 3mm 2mm, clip]{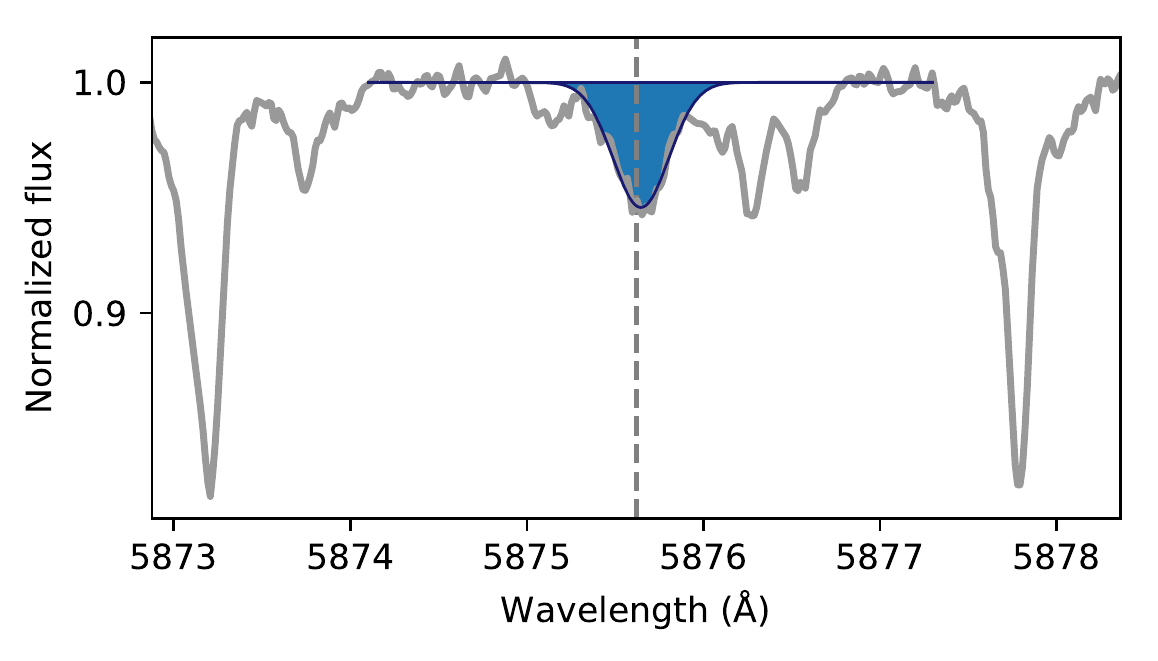}
  \caption{Average stellar spectra around the \lmain{} and \lfive{} regions.
           The stellar spectra are shown by thick gray lines, models are depicted by
           thin blue lines, and integration regions for the equivalent widths are shaded.
           }
  \label{fig:stellar_ew}
\end{figure}

An approximation of the filling factor of dark \hei{} patches can be determined through the equivalent width (EW) ratio of the \lmain{} lines and the optical \lfive{} line \citep{Andretta2017}. To measure these lines, we combined all out-of-transit spectra from all nights to obtain a master spectrum.
The helium \lmain{} lines are located in the wing of the \ion{Si}{i} line at 10827.091~\AA{} (see Fig~\ref{fig:stellar_ew}). As noted by \citet{Andretta2017}, this line is poorly reproduced with a single Voigt profile, so we fitted the line with two superposed Voigt profiles with the same central wavelength. The \lmain{} lines were fitted with Gaussians with fixed relative wavelength but free relative strengths. The \ion{Si}{i} line is found within the stellar rest frame velocity to an accuracy of 140~\ms{}, but for the \lmain{} lines we find a redshift of 0.94~\kms{}. The main component has an equivalent width of 323~m\AA{} and the minor component of 52~m\AA{}, resulting in an EW ratio of 6.2. For the optical \lfive{} line, we also fitted a Gaussian profile and 
derived an equivalent width of 21.5~m\AA{} along with a redshift of 1.3~\kms{}. In the fit, we neglected some minor line blends identified by \citet{Andretta2017}.
The resulting radial velocity shift is similar in the optical and infrared helium lines. 

According to the EWs of the helium lines, \hdo{} is located above the theoretical curve for a helium spot filling factor of 100\% adopted by \citep[][see their Fig.~10]{Andretta2017}, which can likely be attributed to insufficient stellar atmosphere models.
Among the stellar sample studied by \citet{Andretta2017}, $\epsilon$~Eri shows properties comparable to \hdo. In particular, this active K dwarf shows values of 258~m\AA{} and 51~m\AA{} for the EWs of the stellar infrared triplet
components and 18.1~m\AA{} for the optical line along with an X-ray luminosity of $2.1\times10^{28}$~\ergs. For $\epsilon$~Eri, \citet{Andretta2017} derived a minimum filling factor of $59$\% for dark \hei{} patches. By analogy, we adopt a high helium spot filling factor of $75$\% for our pseudo-signal analysis in \hdo.
Such a large filling factor is also consistent  with other aspects of our data, viz,  the absence of both significant inter-transit variability in the stellar \hei{} line and detectable spot crossing events in our data (Sects.~\ref{Sect:APP_absdepth} and \ref{sec:TempEvol}).

We reconstructed the average stellar spectrum shown in the top panel of Fig.~\ref{fig:tsp} by assuming that the complete stellar \lmain{} absorption EW is produced by 75\% dark patches on the stellar surface. For the remaining 25\% of  bright regions we assumed negligible \hei{} absorption. This procedure provides estimates for the spectra of bright and dark patches on the stellar surface. 

\subsubsection{Quantifying the pseudo-signal}

To compute model spectral time series, we adopted a discretized stellar surface, rigidly rotating with a projected rotation velocity, $v\sin i$, of 3.5~\kms.
In the planetary rest frame, the stellar surface is Doppler shifted, most pronouncedly during ingress and egress. The  amount of this shift depends on the relative motion of the star and the planet and the motion of the rotating stellar surface elements.
We used the two-component stellar surface model to calculate spectral time series with the planet occulting only \hei{} dark or bright patches. The occulted spectrum was removed from the in-transit spectrum and division by the out-of-transit spectrum provided the model residual spectra, which were averaged for the ingress, mid-transit, and egress phases to obtain estimates for the pseudo-signals.

\begin{figure}[t]
  \centering
  \includegraphics[width=\hsize, trim = 2.5mm 2.5mm 3mm 2mm, clip]{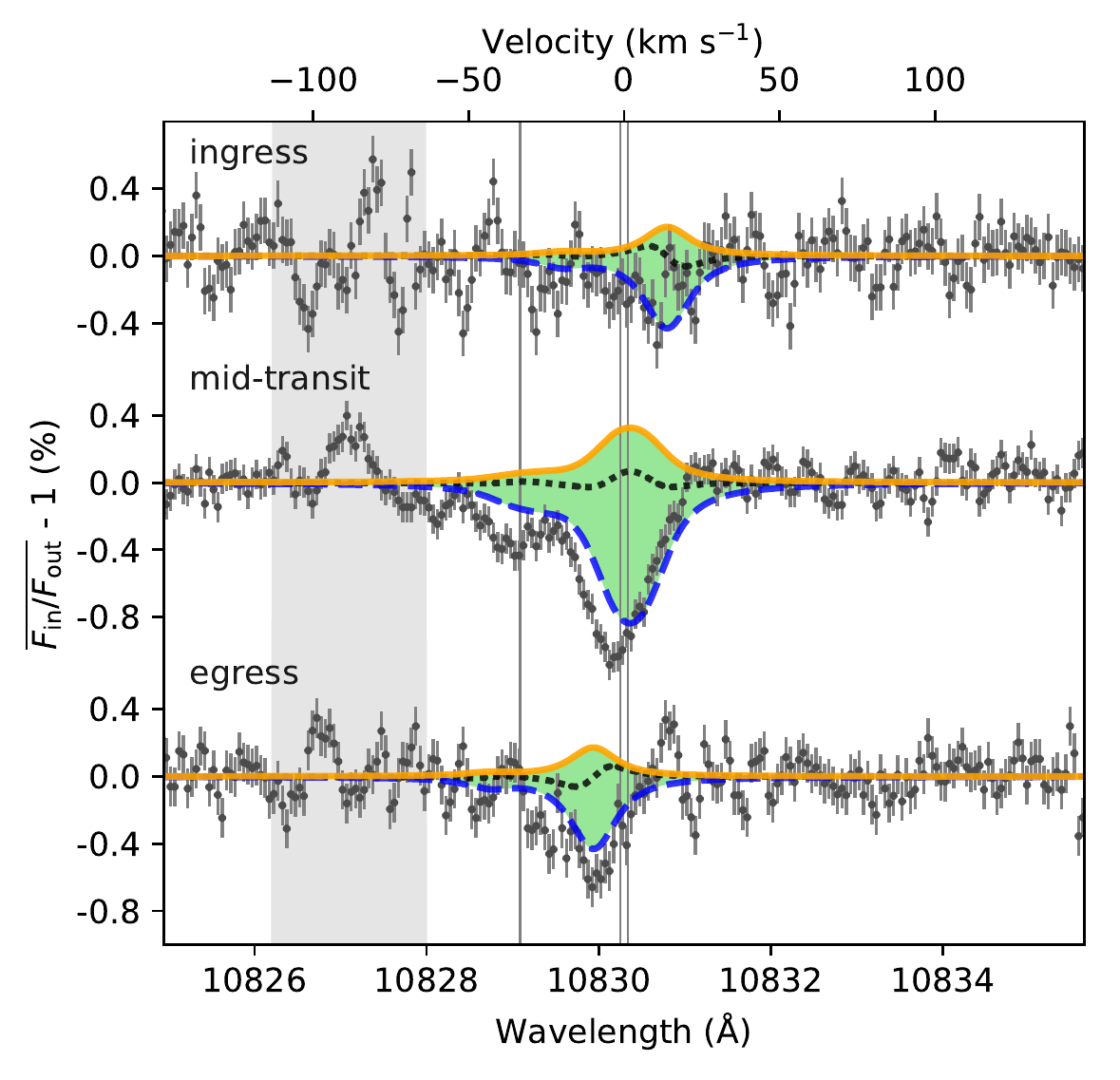}
  \caption{Pseudo-signals created by stellar surface heterogeneity shown in
           comparison to the observed
           ingress, mid-transit, and egress signals.
           The transit over dark patches creates pseudo-emission (orange solid lines),
           and a transit over bright patches creates a pseudo-absorption signal (blue dashed lines).
           Between the two extremes is the possible range of pseudo-signal
           contribution to the observed signal (green shaded areas).
           The gray dotted line indicates the impact of the RME
           neglecting stellar surface inhomogeneity.
           }
  \label{fig:mock}
\end{figure} 

The resulting pseudo-signals are shown in Fig.~\ref{fig:mock}. The pseudo-absorption signal is on a par in strength with the observed signal at all phases, but several features of the observations are not reproduced: 
($i$) the model predicts too little absorption during mid-transit,
10.6~m\AA{} compared to the observed $12.7\pm0.6$~m\AA{}; ($ii$) the line ratio of the pseudo absorption reflects that of the host star of 6.2, which is inconsistent with the observed value of $2.8\pm0.2$; ($iii$) the pseudo-absorption line is redshifted by 4.5~\kms{} with respect to the observations at mid-transit; ($iv$) the model predicts symmetric red- and blueshifted absorption at ingress and egress, which is not observed.
Furthermore, such strong pseudo-absorption signals need a very special geometric configuration as the  planet traverses about 10\% of the stellar disk during the mid-transit phase and these signals would have to cover only bright stellar surface patches that also cover only 25\% of the stellar disk. Moreover, this configuration would have to be very similar for all three transits, which  were observed more than one year apart.

These shortcomings make it unlikely that the observed transit signals are exclusively pseudo-signals. However, the observed radial velocities of the absorption line during ingress  and egress, place the planetary signal close to the stellar rest frame, which hinders a clear distinction between planetary absorption and stellar pseudo-signals.
At any rate, an active star can produce strong features in the residual \hei{} spectra with an amplitude anywhere between the limiting cases presented in Fig.~\ref{fig:mock}.

A special case of the above assumptions that is very likely to interfere with the planetary signal are center-to-limb variations in the stellar \lmain{} lines. The Sun shows limb-darkening in the \lmain{} lines, i.e., the chromospheric \hei{} absorption profile is deeper at the solar rim than in the center of the disk \citep{deJager1966}. This modulation produces phase-dependent pseudo-signals.  
If the \lmain{} lines of \hdo{} behave as those of much less active Sun, we would expect pseudo-emission signals during ingress and egress, and a pseudo-absorption signal during the mid-transit phase. Such a stellar contribution to the supposed planetary absorption signal would place the radial velocity of the absorption signal between the planetary and stellar rest frames during the mid-transit phase. This may have contributed to deflections from the planetary rest frame that are suggested in Fig.~\ref{fig:tsp}, but since they only affected the first two nights, where the instrumental stability was suboptimal, we refrain from further interpretations.

In this context, we also investigated the impact of the RME in the \lmain{} lines. Assuming that the \hei{} lines are homogeneous on the stellar disk, the RME causes signals with an amplitude of less than 0.07\% in the residual spectra (see Fig.~\ref{fig:mock}). The effect is negligible during the mid-transit phase and it is smaller than the possible impact of center-to-limb variations during the ingress and egress phases. Therefore, we do not include the RME in the following analysis.

\subsection{Transmission spectrum modeling}\label{SectModel}

We now assume the other extreme, namely a homogeneous stellar \hei{} disk that causes no pseudo-signals. Neglecting  the structure of the planetary atmosphere as well, we modeled the wavelength-dependent transmission, $T(\lambda)$, with a single absorption component:
\begin{align}
 T(\lambda) = (1-f)+f\exp{[-N^*_{\ion{He}{i}}\sigma(\lambda)]} \, .
\label{eq:transmission}
\end{align}
Here, $f$ denotes the fraction of the stellar disk covered by the \ion{He}{i} cloud, $N^*_{\ion{He}{i}}$  the column density of excited helium, and $\sigma(\lambda)$ the wavelength-dependent absorption cross section, which we parameterized by three Gaussians with their central wavelengths and oscillator strengths fixed to the known values from atomic physics \citep[see National Institute of Standards and Technology, NIST;][]{Drake2006}. The free parameters in our model are therefore the covering fraction, the \ion{He}{i}~column density in the triplet state, and a common velocity shift and line width.

Since the covering fraction and the column density are highly correlated, we adopted two opposing extreme values of $f=1.1$\% and 20\% in our analysis. Assuming that the material is distributed in an annulus surrounding the opaque planet body, the covering fractions correspond to an atmosphere extending to 1.2 and 3.0~\Rpl{}\footnote{The atmosphere starts at 1~\Rpl{} and has a radial extent of 0.2 or 2.0~\Rpl{}.} The former represents the minimum atmospheric extent that can produce the observed 1\% absorption signal and the latter is the effective Roche lobe radius of \hdob{} \citep[Eq.~2 of ][]{Eggleton1983}. In the following we refer to them as the compact and the extended assumption.
For the ingress and egress phases, we derive the average covering fraction for an atmospheric ring at the exact observing times. To that end, we use analytic light curve models \citep{Mandel2002} and subtract the solid body light curve from that with an opaque atmosphere with the given expansion. 

Adopting uniform priors for all parameters, we explored the posterior probability distributions with the Markov chain Monte Carlo technique\footnote{See the PyAstronomy wrapper for PyMC (\url{https://github.com/pymc-devs/pymc})} (MCMC). The chains were run over $10^5$ steps with a burn-in of $10^4$ steps. Our results for the ingress, mid-transit, and egress phase are summarized in Table~\ref{TabFit}.
There we also provide $\chi^2$ values and p-values for the null hypothesis that the maximum likelihood solution is true. Our maximum likelihood models are shown in Fig.~\ref{fig:mean}. 

From our spectral modeling, we find average radial velocities of $6.5\pm3.1$~\kms{} for ingress, $-3.5\pm0.4$~\kms{} for mid-transit, and $-12.6\pm1.0$~\kms{} for egress, independent of the adopted atmospheric extent. We stress once more that these velocities measurements are potentially affected by stellar pseudo-signals (Sect.~\ref{SectSpots}).
At mid-transit, the ratio of the \lmain{} to $\uplambda10829$~\AA{} components deviates from the optically thin ratio of 8 \citep[NIST,][]{Drake2006}. This can be explained by sufficiently large column densities, because saturation in the stronger component of the triplet increases the relative depth of the weaker component when the optically thin approximation breaks down. The observed ratio of $2.8\pm0.2$ corresponds to an optical depth of about 3.2 in the main component \citep{deJager1966}.
While the signal during the ingress and egress phases is equivalently reproduced using either assumptions (see Table~\ref{TabFit}),  the compact case provides a superior approximation for the mid-transit phase. In particular, the extended assumption yields a larger than observed line ratio of 7.9 and a total equivalent width of 11.4~m\AA{}, whereas the compact assumption results in an equivalent width of 12.0~m\AA{} and a line ratio of 4.6, which better reproduces the observed \lminor{} absorption. Formally, this is reflected by a decrease from 1.34 to 1.05 in the reduced $\chi^2$ statistics (see Table~\ref{TabFit}) and a p-value of 0.3 for the compact assumption, which provides no evidence against the null. Finally, we note that the line ratio is also not fully recovered under the compact assumption  because the main component becomes too broad before the depth of the minor component is reproduced. Nevertheless, the mid-transit line ratio strongly favors a small covering fraction, which corresponds to a compact atmosphere.

While we do not fit a proper atmosphere model here, the derived \hei{} triplet state column density is to be understood as an effective value, which can be compared to theoretical models.
From the evaporation model of \citet{Oklopcic2018} for \hdtb{}, we derived a weighted mean absorption height of 1.6~\Rpl{} with a column of about $7.9\times10^{11}$~cm$^{-2}$. We used $N^*_{\ion{He}{i}} R_{\rm p}$ as weights, which accounts for the increasing geometric weight of higher atmospheric layers.
Although the model is for the \hdt{} system, the column density lies between the values derived for our compact and extended cases considered above, which shows that absorption in a planetary atmosphere is a viable origin of the observed signals.
It is not unlikely that evaporation models like those of \citet{Oklopcic2018} can reproduce the mid-transit signal.
Our finding that a compact atmosphere better reproduces the data than a compact atmosphere does not exclude that the atmosphere of \hdob{} is evaporating, but could simply mean that the helium triplet state is not significantly populated in atmospheric layers above 0.2~\Rpl.

In the end, neither of our fits is fully satisfactory, particularly in the region of the weaker triplet component at $\uplambda10829$~\AA{}. We attribute this to shortcomings of the adopted model. A more comprehensive model should consider both pseudo-signals and dedicated atmospheric transmission models.

\begin{table*}
  \caption{MCMC results with 1$\sigma$ errors.}
  \label{TabFit}
  \centering
  \begin{tabular}{l c c c c c c c}
    \hline\hline\vspace{-7pt}\\
     Phase          & $f$\tablefootmark{a} & $N^*_{\ion{He}{i}}$ & $b$\tablefootmark{b}  & Rad. vel. & Line ratio  & $\chi^2_{\rm red}$ & p-value \\
                    & (\%) & ($10^{11}$ cm$^{-2}$)  & (\kms{})        & (\kms{})   &                         & (174 DOF\tablefootmark{c}) & \\
    \vspace{-9pt}\\\hline\vspace{-7pt}\\
    \multicolumn{8}{c}{\it extended atmosphere}\\\vspace{-9pt}\\ 
     T$_1$ - T$_2$  & 9.5       & $0.62\pm 0.11$       &  $15.0\pm 5.0$  & \hphantom{$-1$}$6.4\pm 3.2$ &     & 1.15 & 0.09 \\
     T$_2$ - T$_3$  & 20\hphantom{1.0}& $1.04\pm 0.03$ &  $15.1\pm 0.6$  &   \hphantom{1}$-3.7\pm 0.4$ & 7.9 & 1.34 & 0.002 \\
     T$_3$ - T$_4$  & 9.5       & $1.07\pm 0.07$       &  $13.7\pm 1.0$  &              $-12.2\pm 0.9$ &     & 1.17 & 0.06 \\
    \vspace{-9pt}\\\multicolumn{8}{c}{\it compact atmosphere}\\ \vspace{-9pt}\\                                      
     T$_1$ - T$_2$  & 0.58      & $12.9\pm 2.2$        &  $13.0\pm 4.9$  & \hphantom{$-1$}$6.6\pm 3.0$ &     & 1.14 & 0.10 \\
     T$_2$ - T$_3$  & 1.1\hphantom{1} & $35.7\pm 2.0$        &  $11.3\pm 0.5$  &   \hphantom{1}$-3.3\pm 0.4$ & 4.6 & 1.05 & 0.30 \\
     T$_3$ - T$_4$  & 0.58      & $29.8\pm 3.7$        &  $11.7\pm 0.9$  &              $-13.0\pm 0.9$ &     & 1.17 & 0.07 \\
    \vspace{-9pt}\\\hline
  \end{tabular}
  \tablefoot{
             \tablefoottext{a}{Adopted covering fraction of the stellar disk covered by the planetary atmosphere.}
             \tablefoottext{b}{Doppler parameter given by $\sqrt{2}$ times the velocity dispersion of the Gaussians.}
             \tablefoottext{c}{DOF: degrees of freedom.}
            }
\end{table*}

\subsection{Velocity structure and ingress-egress asymmetry}\label{SectVelo}

The observed absorption is about twice as strong at egress as at  the ingress phase, and it exhibits velocity shifts from the planetary rest frame at all phases.
If \hdob{} was tidally locked and its atmosphere rotated as a solid body, we would expect symmetric radial velocities of the signals at ingress and egress of $\pm$\,3.5~\kms{} if the absorption arises in atmospheric layers at a height of 0.2~\Rpl{}.
However, the observed shifts are asymmetric and significantly exceed this value at egress. We see two plausible hypotheses explaining the observed features, viz,  atmospheric circulation in a dense helium atmosphere or an additional upper, low-density, and asymmetrically expanding atmosphere.

\subsubsection{Atmospheric circulation}

Hydrodynamic models of the irradiated atmospheres of synchronized hot Jupiters predict rather complex circulation patterns. For the specific case of \hdob, the atmospheric circulation model by \citet{Showman2013}, which covers a pressure range in the atmosphere from 2 to $200\times10^{-6}$~bar, predicts the presence of a superrotating equatorial jet, where the bulk velocity increases with height in the atmosphere. Additionally, the model shows a general day-to-night side flow across the poles in high altitude layers, which is the main cause for the predicted net blueshift of around $-3$~\kms{} of molecular absorption signals in transmission spectra \citep[see Fig.~12 of][]{Showman2013}.

Molecular absorption of CO and H$_2$O indicate such a net blueshift \citep[$-1.7\pm1.2$~\kms{} and $-1.6^{+3.2}_{-2.7}$~\kms{}, respectively;][]{Brogi2016, Brogi2018}.
These signals are sensitive to pressure levels between 0.1 and $10^{-6}$~bar. In contrast, ground-based high-resolution transmission spectroscopy in the sodium lines is sensitive to the lower pressure levels ($<$\,$10^{-6}$~bar) that are reached in the lower planetary thermosphere \citep{Pino2018}. Despite their origin in higher atmospheric layers, the sodium absorption signals of \hdob{}  indicate a net blueshift of $-1.9\pm0.7$~\kms{} during mid-transit \citep{Louden2015}.
The  helium absorption presented here likely probes even lower pressure levels in the planetary atmosphere. Particularly, the peak \hei{} column density occurs at a pressure level of around $10^{-9}$~bar in the evaporation models of \citet{Oklopcic2018}. 
Nevertheless, our mid-transit \hei{} absorption signal also exhibits a significant blueshift of $-3.5\pm0.4$~\kms, which is consistent with the previous results for the lower atmosphere.

At ingress and egress, the sodium signal indicates radial velocities of $+2.3^{+1.3}_{-1.5}$~\kms{} and $-5.3^{+1.0}_{-1.4}$~\kms{}, which are believed to be caused by an equatorial superrotating jet. At ingress the absorption signal is dominated by the leading limb where superrotating material moves away from the observer, which would explain the observed redshift. At egress the situation is reversed.
The amplitude of the observed bulk velocities in the sodium signal are about a factor of 2.5 smaller than our results, but they do exhibit the same red- to blueshifted asymmetry \citep{Louden2015}. While our \hei{} ingress and egress velocities are also larger than those predicted by the circulation models of \citet{Showman2013}, 
the observed pressure levels are clearly beyond the modeled atmospheric range.

If the advection timescale is comparable to the de-excitation timescale of the \hei{} triplet state, the equatorial jet transports ground level helium atoms from the night side to the leading atmospheric limb and excited helium atoms from the dayside to the trailing limb. This naturally causes stronger \lmain{} absorption at egress compared to the ingress phase. If we approximate the advection timescale by dividing the observed average ingress/egress radial velocities by the planetary radius, we derive a value of $10^{-4}$~s$^{-1}$. This is on the same order as the radiative transition rate to the ground level $A_{31} = 1.272\times10^{-4}$~s$^{-1}$ \citep{Drake2006}. Depending on the local conditions other processes could be faster in depopulating the metastable state, but it seems reasonable that the superrotating jet can also cause strong egress absorption through advection of excited helium atoms from the dayside.

Overall, the observations are consistent with previous observations and with the models of \citet{Showman2013} if the equatorial superrotating jet continues to exhibit increasing bulk velocities in higher atmospheric layers.

\subsubsection{Asymmetric expanding atmosphere}

The observed ingress signal is significant only at the 2$\sigma$ level. If we attribute the redshift of the ingress signal to rednoise or another source unrelated to the planetary atmosphere, we are left with slightly blueshifted absorption at mid-transit and a larger blueshift at egress. The mid-transit signal is consistent with being caused by dense material in a compact atmosphere as shown in Sect.~\ref{SectModel}, but the density of the material that dominates the egress signal is not confined by our data. 

The blueshifted radial velocities could be explained by material that evaporates from the planet and is subsequently being pushed backward, perhaps as a result of the stellar wind pressure creating an asymmetrically expanding atmosphere. The mechanism would be similar to the Type~I interaction studied by \citet{Matsakos2015}.
In this case, the compact atmosphere that is observed during mid-transit causes only a small contribution to the observed egress signal similar to that during ingress. If the asymmetrically expanding atmosphere trails the planet, it would still cover the stellar disk at egress and dominate the observed absorption at this phase. The observed egress radial velocity would then be a measure of the bulk radial velocity of the evaporating material streaming away from the planet with velocity components pointing out of the system and in the reverse direction of orbit motion.
At mid-transit the trailing material would be superposed onto that of the compact atmosphere causing the observed blueshift.

The nondetection of post-transit absorption constrains the (projected) extent of the hypothesized distribution of trailing material observed by means of \hei{} in the triplet state. The lack of post-transit absorption could be explained by a tail structure that is nearly aligned with the star-planet axis. In fact, the 3D model of \citet{Spake2018} for WASP-107\,b demonstrates that radiation pressure can create such a tail. However, we do not detect the strong blueshifts of the tail material predicted by the authors.
A better explanation comes from the the spherical evaporation models of \citet{Oklopcic2018}. The average absorption height for \hdtb{} was 1.6~\Rpl{} (Sect.~\ref{SectModel}), and the triplet state density quickly decreased at higher atmospheric levels. If this characteristic height also applies to an asymmetric extended atmosphere, it is consistent with our nondetection of post-transit absorption because excited helium atoms are not expected at large distances from the planet.

We therefore find the observations consistent with signals from a superposition of a dense, symmetric helium atmosphere and an asymmetrically expanding component that streams away from the planet and slightly trails it.

\subsection{Comparison to \lya{} absorption}\label{SectLya}

\lya{} observations have revealed variable absorption during the transit of \hdob{} at high blueshifts between $-$230 and $-$140~\kms{} \citep{Lecavelier2012, Bourrier2013}.
Applying the same bootstrap method as in Sect.~\ref{Sect:APP_absdepth}, we determine a mean absorption level of $-0.017\pm0.018$\,\% in this range, which is consistent with no \hei{} absorption.

In the optically thin limit, the absorption EW is proportional to the cross section , $\sigma$, times the column density
\begin{equation}
  {\rm EW} = N\sigma = \frac{N \lambda^2 \pi e^2 f}{m_{\rm e} c} \sim \lambda^2 f,
\end{equation}
where $\lambda$ is the central wavelength and $f$ the oscillator strength. The two blended lines of the \lmain{} triplet are by a factor of 90 more strongly absorbed than the hydrogen \lya{} line \citep[NIST,][]{Drake2006}. The relative abundance of neutral hydrogen to that of helium in the metastable triplet state is about $10^5$ in the simulations of \citet{Oklopcic2018}. Thus, the \lya{} line traces different atmospheric layers that can be a factor $10^3$ more rarefied.

We note that \citet{Bourrier2013B} interpreted the \lya{} absorption signal of \hdob{} in terms of the presence of energetic neutrals created through charge exchange of stellar wind protons with neutrals in the planetary atmosphere. Energetic neutrals comprise a different population and their density depends on parameters like the stellar wind density and velocity. Currently, we cannot assess whether charge exchange can also create substantial amounts of helium in the excited triplet state.
The variability of the \lya{} signal, with absorption detected in only about half of the observations \citep{Bourrier2013}, is in contrast with the stability of the helium absorption signal. This fosters the picture that the two signals arise in populations that are decoupled to a larger degree, i.e., that the \lya{} absorption at large blueshifts originates from the interaction of the stellar wind with the planetary atmosphere, and that the helium absorption occurs in the thermosphere closer to the planetary body.

%_______________________________________________________________________________
%_______________________________________________________________________________
\section{Conclusions}\label{SectConclusions}
%_______________________________________________________________________________
%_______________________________________________________________________________
We present the detection of spectrally resolved \hei{} absorption signals in the near-infrared during three individual transit observations of \hdob{} with CARMENES.
The mid-transit signal in the \lmain{} main component has a depth of $0.88\pm0.04$\,\%. It exhibits a net blueshift of $-3.5\pm0.4$~\kms, and shows no detectable variation in strength between the three transits. The ingress and egress signals show red- and blueshifts of $+6.5\pm3.1$~\kms{} and $-12.6\pm1.0$~\kms{}, respectively.

Our analysis reveals that pseudo-signals induced by the stellar surface structure in the \lmain{} lines might interfere with the atmospheric signal of the planet, but do not reproduce all features of the data. We consider it unlikely that pseudo-signals can exclusively explain the transit signal. In the worst-case, we estimate that they could account for up to 80\% of the detected signal strength. Additionally, pseudo-signals might also affect the measured radial velocities at all phases.

When interpreted in terms of planetary atmospheric absorption, a compact atmosphere is favored with an extent of 0.2 planetary radii, which is easily contained within the planetary Roche lobe. This is consistent with the lack of both clear pre- or post-transit absorption and \hei{} at radial velocities exceeding the planetary escape velocity. We discuss two hypotheses to explain the observed radial velocity signature, namely, atmospheric circulation in the upper planetary atmosphere and an asymmetric extended atmosphere of evaporating material.

Atmospheric circulation with equatorial superrotation has been indicated in observations of lower atmospheric layers, which it is also predicted by models, and here we propose that it might persist throughout the higher layers responsible for the \hei{} absorption. The superrotation hypothesis hinges on the signals and radial velocity shifts during the ingress and egress phases. While we consider the latter significant, the result for the ingress phase is more uncertain.  
If we attribute the observed redshift during ingress to an unrelated source, such as red-noise interference or unaccounted for stellar effects, the case for atmospheric superrotation wanes. In this case, the hypothesis of an asymmetrically evaporating atmosphere that accounts for the blueshifted egress signal becomes more attractive. Indeed, models of planetary evaporation predict such structures and \lya{} observations have demonstrated the existence of material at large blueshifts exceeding $-$140~\kms. However, the lower radial velocity of the helium signal shows that we observe a different region of the planetary atmosphere. 

Our analysis shows that transit spectroscopy of the \hei{} line is a highly promising tool for the study of planetary atmospheric physics. Although, the atmosphere of \hdob{} is almost certainly escaping from the planet, as evidenced by the \lya{} observations, it remains uncertain how well the mass-loss rate can be determined from \lmain{} absorption and to what degree the atmospheric absorption can be distinguished from stellar pseudo-signals. 
Detailed modeling is needed to investigate the physical plausibility of the two sketched interpretations, and only new observations will allow us to distinguish between them by a confirmation or rejection of the ingress signal.

\begin{acknowledgements}
CARMENES is an instrument for the Centro Astron\'omico Hispano-Alem\'an de
Calar Alto (CAHA, Almer\'{\i}a, Spain). 
CARMENES is funded by the German Max-Planck-Gesellschaft (MPG), 
the Spanish Consejo Superior de Investigaciones Cient\'{\i}ficas (CSIC),
the European Union through FEDER/ERF FICTS-2011-02 funds, 
and the members of the CARMENES Consortium 
(Max-Planck-Institut f\"ur Astronomie,
Instituto de Astrof\'{\i}sica de Andaluc\'{\i}a,
Landessternwarte K\"onigstuhl,
Institut de Ci\`encies de l'Espai,
Insitut f\"ur Astrophysik G\"ottingen,
Universidad Complutense de Madrid,
Th\"uringer Landessternwarte Tautenburg,
Instituto de Astrof\'{\i}sica de Canarias,
Hamburger Sternwarte,
Centro de Astrobiolog\'{\i}a and
Centro Astron\'omico Hispano-Alem\'an), 
with additional contributions 
by the Spanish Ministerio de Ciencia, Innovaci\'on y Universidades  
through projects ESP2013-48391-C4-1-R, ESP2014-54062-R,  
ESP2014-54362-P, ESP2014-57495-C2-2-R, AYA2015-69350-C3-2-P,  
AYA2016-79425-C3-1/2/3-P, ESP2016 76076-R, ESP2016-80435-C2-1-R,  
ESP2017-87143-R, and AYA2018-84089;
the German Science Foundation through the Major Research Instrumentation 
Programme and DFG Research Unit FOR2544 ``Blue Planets around Red Stars''; 
the Klaus Tschira Stiftung; 
the states of Baden-W\"urttemberg and Niedersachsen; 
and by the Junta de Andaluc\'{\i}a.\\
We also acknowledge support from the Deutsche  
Forschungsgemeinschaft through projects SCHM~1032/57-1 and SCH~1382/2-1, the Deutsches  
Zentrum f\"ur Luft- und Raumfahrt through projects 50OR1706 and  
50OR1710, the European Research Council through project No~694513, the  
Fondo Europeo de Desarrollo Regional, and the Generalitat de  
Catalunya/CERCA programme.
This work has made use of data from the European Space Agency (ESA) mission
{\it Gaia} (\url{https://www.cosmos.esa.int/gaia}), processed by the {\it Gaia}
Data Processing and Analysis Consortium (DPAC,
\url{https://www.cosmos.esa.int/web/gaia/dpac/consortium}). Funding for the DPAC
has been provided by national institutions, in particular the institutions
participating in the {\it Gaia} Multilateral Agreement.
Finally, we thank the referee for constructive comments that helped to improve this publication.
\end{acknowledgements}

\bibliographystyle{aa}
\setlength{\bibsep}{0.0pt}
\bibliography{hd189_he10830}

\newcommand{\noop}[1]{}
\begin{thebibliography}{97}
\expandafter\ifx\csname natexlab\endcsname\relax\def\natexlab#1{#1}\fi

\bibitem[{{Agol} {et~al.}(2010){Agol}, {Cowan}, {Knutson}, {Deming}, {Steffen},
  {Henry}, \& {Charbonneau}}]{Agol2010}
{Agol}, E., {Cowan}, N.~B., {Knutson}, H.~A., {et~al.} 2010, \apj, 721, 1861

\bibitem[{{Allart} {et~al.}(2017){Allart}, {Lovis}, {Pino}, {Wyttenbach},
  {Ehrenreich}, \& {Pepe}}]{Allart2017}
{Allart}, R., {Lovis}, C., {Pino}, L., {et~al.} 2017, \aap, 606, A144

\bibitem[{{Andretta} {et~al.}(2017){Andretta}, {Giampapa}, {Covino}, {Reiners},
  \& {Beeck}}]{Andretta2017}
{Andretta}, V., {Giampapa}, M.~S., {Covino}, E., {Reiners}, A., \& {Beeck}, B.
  2017, \apj, 839, 97

\bibitem[{{Astropy Collaboration} {et~al.}(2013){Astropy Collaboration},
  {Robitaille}, {Tollerud}, {Greenfield}, {Droettboom}, {Bray}, {Aldcroft},
  {Davis}, {Ginsburg}, {Price-Whelan}, {Kerzendorf}, {Conley}, {Crighton},
  {Barbary}, {Muna}, {Ferguson}, {Grollier}, {Parikh}, {Nair}, {Unther},
  {Deil}, {Woillez}, {Conseil}, {Kramer}, {Turner}, {Singer}, {Fox}, {Weaver},
  {Zabalza}, {Edwards}, {Azalee Bostroem}, {Burke}, {Casey}, {Crawford},
  {Dencheva}, {Ely}, {Jenness}, {Labrie}, {Lim}, {Pierfederici}, {Pontzen},
  {Ptak}, {Refsdal}, {Servillat}, \& {Streicher}}]{Astropy2013}
{Astropy Collaboration}, {Robitaille}, T.~P., {Tollerud}, E.~J., {et~al.} 2013,
  \aap, 558, A33

\bibitem[{{Avrett} {et~al.}(1994){Avrett}, {Fontenla}, \&
  {Loeser}}]{Avrett1994}
{Avrett}, E.~H., {Fontenla}, J.~M., \& {Loeser}, R. 1994, in IAU Symposium,
  Vol. 154, Infrared Solar Physics, ed. D.~M. {Rabin}, J.~T. {Jefferies}, \&
  C.~{Lindsey}, 35

\bibitem[{{Baliunas} {et~al.}(1995){Baliunas}, {Donahue}, {Soon}, {Horne},
  {Frazer}, {Woodard-Eklund}, {Bradford}, {Rao}, {Wilson}, {Zhang}, {Bennett},
  {Briggs}, {Carroll}, {Duncan}, {Figueroa}, {Lanning}, {Misch}, {Mueller},
  {Noyes}, {Poppe}, {Porter}, {Robinson}, {Russell}, {Shelton}, {Soyumer},
  {Vaughan}, \& {Whitney}}]{Baliunas1995}
{Baliunas}, S.~L., {Donahue}, R.~A., {Soon}, W.~H., {et~al.} 1995, \apj, 438,
  269

\bibitem[{{Ballester} \& {Ben-Jaffel}(2015)}]{Ballester2015}
{Ballester}, G.~E. \& {Ben-Jaffel}, L. 2015, \apj, 804, 116

\bibitem[{{Baluev} {et~al.}(2015){Baluev}, {Sokov}, {Shaidulin}, {Sokova},
  {Jones}, {Tuomi}, {Anglada-Escud{\'e}}, {Benni}, {Colazo}, {Schneiter},
  {D'Angelo}, {Burdanov}, {Fern{\'a}ndez-Laj{\'u}s}, {Ba{\c s}t{\"u}rk},
  {Hentunen}, \& {Shadick}}]{Baluev2015}
{Baluev}, R.~V., {Sokov}, E.~N., {Shaidulin}, V.~S., {et~al.} 2015, \mnras,
  450, 3101

\bibitem[{{Barnes} {et~al.}(2016){Barnes}, {Haswell}, {Staab}, \&
  {Anglada-Escud{\'e}}}]{Barnes2016}
{Barnes}, J.~R., {Haswell}, C.~A., {Staab}, D., \& {Anglada-Escud{\'e}}, G.
  2016, \mnras, 462, 1012

\bibitem[{{Ben-Jaffel} \& {Ballester}(2013)}]{BenJaffel2013}
{Ben-Jaffel}, L. \& {Ballester}, G.~E. 2013, \aap, 553, A52

\bibitem[{{Ben-Jaffel} \& {Sona Hosseini}(2010)}]{BenJaffel2010}
{Ben-Jaffel}, L. \& {Sona Hosseini}, S. 2010, \apj, 709, 1284

\bibitem[{{Birkby} {et~al.}(2013){Birkby}, {de Kok}, {Brogi}, {de Mooij},
  {Schwarz}, {Albrecht}, \& {Snellen}}]{Birkby2013}
{Birkby}, J.~L., {de Kok}, R.~J., {Brogi}, M., {et~al.} 2013, \mnras, 436, L35

\bibitem[{{Bouchy} {et~al.}(2005){Bouchy}, {Udry}, {Mayor}, {Moutou}, {Pont},
  {Iribarne}, {da Silva}, {Ilovaisky}, {Queloz}, {Santos}, {S{\'e}gransan}, \&
  {Zucker}}]{Bouchy2005}
{Bouchy}, F., {Udry}, S., {Mayor}, M., {et~al.} 2005, \aap, 444, L15

\bibitem[{{Bourrier} \& {Lecavelier des Etangs}(2013)}]{Bourrier2013B}
{Bourrier}, V. \& {Lecavelier des Etangs}, A. 2013, \aap, 557, A124

\bibitem[{{Bourrier} {et~al.}(2013){Bourrier}, {Lecavelier des Etangs},
  {Dupuy}, {Ehrenreich}, {Vidal-Madjar}, {H{\'e}brard}, {Ballester},
  {D{\'e}sert}, {Ferlet}, {Sing}, \& {Wheatley}}]{Bourrier2013}
{Bourrier}, V., {Lecavelier des Etangs}, A., {Dupuy}, H., {et~al.} 2013, \aap,
  551, A63

\bibitem[{{Boyajian} {et~al.}(2015){Boyajian}, {von Braun}, {Feiden}, {Huber},
  {Basu}, {Demarque}, {Fischer}, {Schaefer}, {Mann}, {White}, {Maestro},
  {Brewer}, {Lamell}, {Spada}, {L{\'o}pez-Morales}, {Ireland}, {Farrington},
  {van Belle}, {Kane}, {Jones}, {ten Brummelaar}, {Ciardi}, {McAlister},
  {Ridgway}, {Goldfinger}, {Turner}, \& {Sturmann}}]{Boyajian2015}
{Boyajian}, T., {von Braun}, K., {Feiden}, G.~A., {et~al.} 2015, \mnras, 447,
  846

\bibitem[{{Brogi} {et~al.}(2016){Brogi}, {de Kok}, {Albrecht}, {Snellen},
  {Birkby}, \& {Schwarz}}]{Brogi2016}
{Brogi}, M., {de Kok}, R.~J., {Albrecht}, S., {et~al.} 2016, \apj, 817, 106

\bibitem[{{Brogi} {et~al.}(2018){Brogi}, {Giacobbe}, {Guilluy}, {de Kok},
  {Sozzetti}, {Mancini}, \& {Bonomo}}]{Brogi2018}
{Brogi}, M., {Giacobbe}, P., {Guilluy}, G., {et~al.} 2018, \aap, 615, A16

\bibitem[{{Brown} {et~al.}(2018){Brown}, {Vallenari}, {Prusti}, {de Bruijne},
  {Babusiaux}, {Bailer-Jones}, {Biermann}, {Evans}, {Eyer}, {Jansen}, {Jordi},
  {Klioner}, {Lammers}, {Lindegren}, {Luri}, {Mignard}, {Panem}, {Pourbaix},
  {Randich}, {Sartoretti}, {Siddiqui}, {Soubiran}, {van Leeuwen}, {Walton},
  {Arenou}, {Bastian}, {Cropper}, {Drimmel}, {Katz}, {Lattanzi}, {Bakker},
  {Cacciari}, {Casta{\~n}eda}, {Chaoul}, {Cheek}, {De Angeli}, {Fabricius},
  {Guerra}, {Holl}, {Masana}, {Messineo}, {Mowlavi}, {Nienartowicz}, {Panuzzo},
  {Portell}, {Riello}, {Seabroke}, {Tanga}, {Th{\'e}venin}, {Gracia-Abril},
  {Comoretto}, {Garcia-Reinaldos}, {Teyssier}, {Altmann}, {Andrae}, {Audard},
  {Bellas-Velidis}, {Benson}, {Berthier}, {Blomme}, {Burgess}, {Busso},
  {Carry}, {Cellino}, {Clementini}, {Clotet}, {Creevey}, {Davidson}, {De
  Ridder}, {Delchambre}, {Dell'Oro}, {Ducourant},
  {Fern{\'a}ndez-Hern{\'a}ndez}, {Fouesneau}, {Fr{\'e}mat}, {Galluccio},
  {Garc{\'\i}a-Torres}, {Gonz{\'a}lez-N{\'u}{\~n}ez}, {Gonz{\'a}lez- Vidal},
  {Gosset}, {Guy}, {Halbwachs}, {Hambly}, {Harrison}, {Hern{\'a}ndez},
  {Hestroffer}, {Hodgkin}, {Hutton}, {Jasniewicz}, {Jean-Antoine- Piccolo},
  {Jordan}, {Korn}, {Krone- Martins}, {Lanzafame}, {Lebzelter}, {L{\"o}ffler},
  {Manteiga}, {Marrese}, {Mart{\'\i}n-Fleitas}, {Moitinho}, {Mora}, {Muinonen},
  {Osinde}, {Pancino}, {Pauwels}, {Petit}, {Recio-Blanco}, {Richards},
  {Rimoldini}, {Robin}, {Sarro}, {Siopis}, {Smith}, {Sozzetti}, {S{\"u}veges},
  {Torra}, {van Reeven}, {Abbas}, {Abreu Aramburu}, {Accart}, {Aerts},
  {Altavilla}, {{\'A}lvarez}, {Alvarez}, {Alves}, {Anderson}, {Andrei},
  {Anglada Varela}, {Antiche}, {Antoja}, {Arcay}, {Astraatmadja}, {Bach},
  {Baker}, {Balaguer-N{\'u}{\~n}ez}, {Balm}, {Barache}, {Barata}, {Barbato},
  {Barblan}, {Barklem}, {Barrado}, {Barros}, {Barstow}, {Bartholom{\'e}
  Mu{\~n}oz}, {Bassilana}, {Becciani}, {Bellazzini}, {Berihuete}, {Bertone},
  {Bianchi}, {Bienaym{\'e}}, {Blanco-Cuaresma}, {Boch}, {Boeche}, {Bombrun},
  {Borrachero}, {Bossini}, {Bouquillon}, {Bourda}, {Bragaglia}, {Bramante},
  {Breddels}, {Bressan}, {Brouillet}, {Br{\"u}semeister}, {Brugaletta},
  {Bucciarelli}, {Burlacu}, {Busonero}, {Butkevich}, {Buzzi}, {Caffau},
  {Cancelliere}, {Cannizzaro}, {Cantat-Gaudin}, {Carballo}, {Carlucci},
  {Carrasco}, {Casamiquela}, {Castellani}, {Castro-Ginard}, {Charlot},
  {Chemin}, {Chiavassa}, {Cocozza}, {Costigan}, {Cowell}, {Crifo}, {Crosta},
  {Crowley}, {Cuypers}, {Dafonte}, {Damerdji}, {Dapergolas}, {David}, {David},
  {de Laverny}, {De Luise}, {De March}, {de Martino}, {de Souza}, {de Torres},
  {Debosscher}, {del Pozo}, {Delbo}, {Delgado}, {Delgado}, {Di Matteo},
  {Diakite}, {Diener}, {Distefano}, {Dolding}, {Drazinos}, {Dur{\'a}n},
  {Edvardsson}, {Enke}, {Eriksson}, {Esquej}, {Eynard Bontemps}, {Fabre},
  {Fabrizio}, {Faigler}, {Falc{\~a}o}, {Farr{\`a}s Casas}, {Federici},
  {Fedorets}, {Fernique}, {Figueras}, {Filippi}, {Findeisen}, {Fonti},
  {Fraile}, {Fraser}, {Fr{\'e}zouls}, {Gai}, {Galleti}, {Garabato},
  {Garc{\'\i}a-Sedano}, {Garofalo}, {Garralda}, {Gavel}, {Gavras}, {Gerssen},
  {Geyer}, {Giacobbe}, {Gilmore}, {Girona}, {Giuffrida}, {Glass}, {Gomes},
  {Granvik}, {Gueguen}, {Guerrier}, {Guiraud}, {Guti{\'e}rrez-S{\'a}nchez},
  {Haigron}, {Hatzidimitriou}, {Hauser}, {Haywood}, {Heiter}, {Helmi}, {Heu},
  {Hilger}, {Hobbs}, {Hofmann}, {Holland}, {Huckle}, {Hypki}, {Icardi},
  {Jan{\ss}en}, {Jevardat de Fombelle}, {Jonker}, {Juh{\'a}sz}, {Julbe},
  {Karampelas}, {Kewley}, {Klar}, {Kochoska}, {Kohley}, {Kolenberg},
  {Kontizas}, {Kontizas}, {Koposov}, {Kordopatis}, {Kostrzewa-Rutkowska},
  {Koubsky}, {Lambert}, {Lanza}, {Lasne}, {Lavigne}, {Le Fustec}, {Le
  Poncin-Lafitte}, {Lebreton}, {Leccia}, {Leclerc}, {Lecoeur-Taibi},
  {Lenhardt}, {Leroux}, {Liao}, {Licata}, {Lindstr{\o}m}, {Lister}, {Livanou},
  {Lobel}, {L{\'o}pez}, {Managau}, {Mann}, {Mantelet}, {Marchal}, {Marchant},
  {Marconi}, {Marinoni}, {Marschalk{\'o}}, {Marshall}, {Martino}, {Marton},
  {Mary}, {Massari}, {Matijevi{\v{c}}}, {Mazeh}, {McMillan}, {Messina},
  {Michalik}, {Millar}, {Molina}, {Molinaro}, {Moln{\'a}r}, {Montegriffo},
  {Mor}, {Morbidelli}, {Morel}, {Morris}, {Mulone}, {Muraveva}, {Musella},
  {Nelemans}, {Nicastro}, {Noval}, {O'Mullane}, {Ord{\'e}novic},
  {Ord{\'o}{\~n}ez-Blanco}, {Osborne}, {Pagani}, {Pagano}, {Pailler},
  {Palacin}, {Palaversa}, {Panahi}, {Pawlak}, {Piersimoni}, {Pineau}, {Plachy},
  {Plum}, {Poggio}, {Poujoulet}, {Pr{\v{s}}a}, {Pulone}, {Racero}, {Ragaini},
  {Rambaux}, {Ramos-Lerate}, {Regibo}, {Reyl{\'e}}, {Riclet}, {Ripepi}, {Riva},
  {Rivard}, {Rixon}, {Roegiers}, {Roelens}, {Romero-G{\'o}mez}, {Rowell},
  {Royer}, {Ruiz-Dern}, {Sadowski}, {Sagrist{\`a} Sell{\'e}s}, {Sahlmann},
  {Salgado}, {Salguero}, {Sanna}, {Santana- Ros}, {Sarasso}, {Savietto},
  {Schultheis}, {Sciacca}, {Segol}, {Segovia}, {S{\'e}gransan}, {Shih},
  {Siltala}, {Silva}, {Smart}, {Smith}, {Solano}, {Solitro}, {Sordo}, {Soria
  Nieto}, {Souchay}, {Spagna}, {Spoto}, {Stampa}, {Steele},
  {Steidelm{\"u}ller}, {Stephenson}, {Stoev}, {Suess}, {Surdej}, {Szabados},
  {Szegedi-Elek}, {Tapiador}, {Taris}, {Tauran}, {Taylor}, {Teixeira},
  {Terrett}, {Teyssandier}, {Thuillot}, {Titarenko}, {Torra Clotet}, {Turon},
  {Ulla}, {Utrilla}, {Uzzi}, {Vaillant}, {Valentini}, {Valette}, {van Elteren},
  {Van Hemelryck}, {van Leeuwen}, {Vaschetto}, {Vecchiato}, {Veljanoski},
  {Viala}, {Vicente}, {Vogt}, {von Essen}, {Voss}, {Votruba}, {Voutsinas},
  {Walmsley}, {Weiler}, {Wertz}, {Wevers}, {Wyrzykowski}, {Yoldas},
  {{\v{Z}}erjal}, {Ziaeepour}, {Zorec}, {Zschocke}, {Zucker}, {Zurbach}, \&
  {Zwitter}}]{Gaia2018}
{Brown}, A.~G.~A., {Vallenari}, A., {Prusti}, T., {et~al.} 2018, \aap, 616, A1

\bibitem[{{Caballero} {et~al.}(2016){Caballero}, {Gu{\`a}rdia}, {L{\'o}pez del
  Fresno}, {Zechmeister}, {de Juan}, {Alonso-Floriano}, {Amado}, {Colom{\'e}},
  {Cort{\'e}s-Contreras}, {Garc{\'{\i}}a-Piquer}, {Gesa}, {de Guindos},
  {Hagen}, {Helmling}, {Hern{\'a}ndez Casta{\~n}o}, {K{\"u}rster},
  {L{\'o}pez-Santiago}, {Montes}, {Morales Mu{\~n}oz}, {Pavlov}, {Quirrenbach},
  {Reiners}, {Ribas}, {Seifert}, \& {Solano}}]{Caballero2016}
{Caballero}, J.~A., {Gu{\`a}rdia}, J., {L{\'o}pez del Fresno}, M., {et~al.}
  2016, in \procspie, Vol. 9910, Observatory Operations: Strategies, Processes,
  and Systems VI, 99100E

\bibitem[{{Casasayas-Barris} {et~al.}(2017){Casasayas-Barris}, {Palle},
  {Nowak}, {Yan}, {Nortmann}, \& {Murgas}}]{CasasayasBarris2017}
{Casasayas-Barris}, N., {Palle}, E., {Nowak}, G., {et~al.} 2017, \aap, 608,
  A135

\bibitem[{{Cauley} {et~al.}(2017){Cauley}, {Redfield}, \&
  {Jensen}}]{Cauley2017}
{Cauley}, P.~W., {Redfield}, S., \& {Jensen}, A.~G. 2017, \aj, 153, 217

\bibitem[{{Cegla} {et~al.}(2016){Cegla}, {Lovis}, {Bourrier}, {Beeck},
  {Watson}, \& {Pepe}}]{Cegla2016}
{Cegla}, H.~M., {Lovis}, C., {Bourrier}, V., {et~al.} 2016, \aap, 588, A127

\bibitem[{{Claret} \& {Bloemen}(2011)}]{Claret2011}
{Claret}, A. \& {Bloemen}, S. 2011, \aap, 529, A75

\bibitem[{{Czesla} {et~al.}(2015){Czesla}, {Klocov{\'a}}, {Khalafinejad},
  {Wolter}, \& {Schmitt}}]{Czesla2015}
{Czesla}, S., {Klocov{\'a}}, T., {Khalafinejad}, S., {Wolter}, U., \&
  {Schmitt}, J.~H.~M.~M. 2015, \aap, 582, A51

\bibitem[{{de Jager} {et~al.}(1966){de Jager}, {Namba}, \&
  {Neven}}]{deJager1966}
{de Jager}, C., {Namba}, O., \& {Neven}, L. 1966, \bain, 18, 128

\bibitem[{{de Kok} {et~al.}(2013){de Kok}, {Brogi}, {Snellen}, {Birkby},
  {Albrecht}, \& {de Mooij}}]{deKok2013}
{de Kok}, R.~J., {Brogi}, M., {Snellen}, I.~A.~G., {et~al.} 2013, \aap, 554,
  A82

\bibitem[{{Di Gloria} {et~al.}(2015){Di Gloria}, {Snellen}, \&
  {Albrecht}}]{DiGloria2015}
{Di Gloria}, E., {Snellen}, I.~A.~G., \& {Albrecht}, S. 2015, \aap, 580, A84

\bibitem[{{Drake}(2006)}]{Drake2006}
{Drake}, G. 2006, {High Precision Calculations for Helium}, ed. G.~W.~F.
  {Drake}, 199

\bibitem[{{Eggleton}(1983)}]{Eggleton1983}
{Eggleton}, P.~P. 1983, \apj, 268, 368

\bibitem[{{Ehrenreich} {et~al.}(2015){Ehrenreich}, {Bourrier}, {Wheatley},
  {Lecavelier des Etangs}, {H{\'e}brard}, {Udry}, {Bonfils}, {Delfosse},
  {D{\'e}sert}, {Sing}, \& {Vidal-Madjar}}]{Ehrenreich2015}
{Ehrenreich}, D., {Bourrier}, V., {Wheatley}, P.~J., {et~al.} 2015, \nat, 522,
  459

\bibitem[{{Ehrenreich} {et~al.}(2008){Ehrenreich}, {Lecavelier Des Etangs},
  {H{\'e}brard}, {D{\'e}sert}, {Vidal-Madjar}, {McConnell}, {Parkinson},
  {Ballester}, \& {Ferlet}}]{Ehrenreich2008}
{Ehrenreich}, D., {Lecavelier Des Etangs}, A., {H{\'e}brard}, G., {et~al.}
  2008, \aap, 483, 933

\bibitem[{{Fossati} {et~al.}(2010){Fossati}, {Haswell}, {Froning}, {Hebb},
  {Holmes}, {Kolb}, {Helling}, {Carter}, {Wheatley}, {Collier Cameron},
  {Loeillet}, {Pollacco}, {Street}, {Stempels}, {Simpson}, {Udry}, {Joshi},
  {West}, {Skillen}, \& {Wilson}}]{Fossati2010}
{Fossati}, L., {Haswell}, C.~A., {Froning}, C.~S., {et~al.} 2010, \apjl, 714,
  L222

\bibitem[{{Fulton} {et~al.}(2017){Fulton}, {Petigura}, {Howard}, {Isaacson},
  {Marcy}, {Cargile}, {Hebb}, {Weiss}, {Johnson}, {Morton}, {Sinukoff},
  {Crossfield}, \& {Hirsch}}]{Fulton2017}
{Fulton}, B.~J., {Petigura}, E.~A., {Howard}, A.~W., {et~al.} 2017, \aj, 154,
  109

\bibitem[{{Gaia Collaboration} {et~al.}(2016){Gaia Collaboration}, {Prusti},
  {de Bruijne}, {Brown}, {Vallenari}, {Babusiaux}, {Bailer-Jones}, {Bastian},
  {Biermann}, {Evans}, \& et~al.}]{Gaia2016A}
{Gaia Collaboration}, {Prusti}, T., {de Bruijne}, J.~H.~J., {et~al.} 2016,
  \aap, 595, A1

\bibitem[{{Gibson} {et~al.}(2012){Gibson}, {Aigrain}, {Pont}, {Sing},
  {D{\'e}sert}, {Evans}, {Henry}, {Husnoo}, \& {Knutson}}]{Gibson2012}
{Gibson}, N.~P., {Aigrain}, S., {Pont}, F., {et~al.} 2012, \mnras, 422, 753

\bibitem[{{Harvey} \& {Livingston}(1994)}]{Harvey1994}
{Harvey}, J.~W. \& {Livingston}, W.~C. 1994, in IAU Symposium, Vol. 154,
  Infrared Solar Physics, ed. D.~M. {Rabin}, J.~T. {Jefferies}, \&
  C.~{Lindsey}, 59

\bibitem[{{Haswell} {et~al.}(2012){Haswell}, {Fossati}, {Ayres}, {France},
  {Froning}, {Holmes}, {Kolb}, {Busuttil}, {Street}, {Hebb}, {Collier Cameron},
  {Enoch}, {Burwitz}, {Rodriguez}, {West}, {Pollacco}, {Wheatley}, \&
  {Carter}}]{Haswell2012}
{Haswell}, C.~A., {Fossati}, L., {Ayres}, T., {et~al.} 2012, \apj, 760, 79

\bibitem[{{Henry} \& {Winn}(2008)}]{Henry2008}
{Henry}, G.~W. \& {Winn}, J.~N. 2008, \aj, 135, 68

\bibitem[{{Huitson} {et~al.}(2012){Huitson}, {Sing}, {Vidal-Madjar},
  {Ballester}, {Lecavelier des Etangs}, {D{\'e}sert}, \& {Pont}}]{Huitson2012}
{Huitson}, C.~M., {Sing}, D.~K., {Vidal-Madjar}, A., {et~al.} 2012, \mnras,
  422, 2477

\bibitem[{{H{\"u}nsch} {et~al.}(1999){H{\"u}nsch}, {Schmitt}, {Sterzik}, \&
  {Voges}}]{Huensch1999}
{H{\"u}nsch}, M., {Schmitt}, J.~H.~M.~M., {Sterzik}, M.~F., \& {Voges}, W.
  1999, \aaps, 135, 319

\bibitem[{{Husser} {et~al.}(2013){Husser}, {Wende-von Berg}, {Dreizler},
  {Homeier}, {Reiners}, {Barman}, \& {Hauschildt}}]{Husser2013}
{Husser}, T.-O., {Wende-von Berg}, S., {Dreizler}, S., {et~al.} 2013, \aap,
  553, A6

\bibitem[{{Indriolo} {et~al.}(2009){Indriolo}, {Hobbs}, {Hinkle}, \&
  {McCall}}]{Indriolo2009}
{Indriolo}, N., {Hobbs}, L.~M., {Hinkle}, K.~H., \& {McCall}, B.~J. 2009, \apj,
  703, 2131

\bibitem[{{Jensen} {et~al.}(2012){Jensen}, {Redfield}, {Endl}, {Cochran},
  {Koesterke}, \& {Barman}}]{Jensen2012}
{Jensen}, A.~G., {Redfield}, S., {Endl}, M., {et~al.} 2012, \apj, 751, 86

\bibitem[{{Kausch} {et~al.}(2015){Kausch}, {Noll}, {Smette}, {Kimeswenger},
  {Barden}, {Szyszka}, {Jones}, {Sana}, {Horst}, \& {Kerber}}]{Kausch2015}
{Kausch}, W., {Noll}, S., {Smette}, A., {et~al.} 2015, \aap, 576, A78

\bibitem[{{Khalafinejad} {et~al.}(2017){Khalafinejad}, {von Essen},
  {Hoeijmakers}, {Zhou}, {Klocov{\'a}}, {Schmitt}, {Dreizler}, {Lopez-Morales},
  {Husser}, {Schmidt}, \& {Collet}}]{Sara2017}
{Khalafinejad}, S., {von Essen}, C., {Hoeijmakers}, H.~J., {et~al.} 2017, \aap,
  598, A131

\bibitem[{{Klocov{\'a}} {et~al.}(2017){Klocov{\'a}}, {Czesla}, {Khalafinejad},
  {Wolter}, \& {Schmitt}}]{Klocova2017}
{Klocov{\'a}}, T., {Czesla}, S., {Khalafinejad}, S., {Wolter}, U., \&
  {Schmitt}, J.~H.~M.~M. 2017, \aap, 607, A66

\bibitem[{{Knutson} {et~al.}(2010){Knutson}, {Howard}, \&
  {Isaacson}}]{Knutson2010}
{Knutson}, H.~A., {Howard}, A.~W., \& {Isaacson}, H. 2010, \apj, 720, 1569

\bibitem[{{Kulow} {et~al.}(2014){Kulow}, {France}, {Linsky}, \&
  {Loyd}}]{Kulow2014}
{Kulow}, J.~R., {France}, K., {Linsky}, J., \& {Loyd}, R.~O.~P. 2014, \apj,
  786, 132

\bibitem[{{Lammer} {et~al.}(2003){Lammer}, {Selsis}, {Ribas}, {Guinan},
  {Bauer}, \& {Weiss}}]{Lammer2003}
{Lammer}, H., {Selsis}, F., {Ribas}, I., {et~al.} 2003, \apjl, 598, L121

\bibitem[{{Lavie} {et~al.}(2017){Lavie}, {Ehrenreich}, {Bourrier}, {Lecavelier
  des Etangs}, {Vidal-Madjar}, {Delfosse}, {Gracia Berna}, {Heng}, {Thomas},
  {Udry}, \& {Wheatley}}]{Lavie2017}
{Lavie}, B., {Ehrenreich}, D., {Bourrier}, V., {et~al.} 2017, \aap, 605, L7

\bibitem[{{Lecavelier des Etangs} {et~al.}(2012){Lecavelier des Etangs},
  {Bourrier}, {Wheatley}, {Dupuy}, {Ehrenreich}, {Vidal-Madjar}, {H{\'e}brard},
  {Ballester}, {D{\'e}sert}, {Ferlet}, \& {Sing}}]{Lecavelier2012}
{Lecavelier des Etangs}, A., {Bourrier}, V., {Wheatley}, P.~J., {et~al.} 2012,
  \aap, 543, L4

\bibitem[{{Lecavelier des Etangs} {et~al.}(2010){Lecavelier des Etangs},
  {Ehrenreich}, {Vidal-Madjar}, {Ballester}, {D{\'e}sert}, {Ferlet},
  {H{\'e}brard}, {Sing}, {Tchakoumegni}, \& {Udry}}]{Lecavelier2010}
{Lecavelier des Etangs}, A., {Ehrenreich}, D., {Vidal-Madjar}, A., {et~al.}
  2010, \aap, 514, A72

\bibitem[{{Lecavelier Des Etangs} {et~al.}(2008){Lecavelier Des Etangs},
  {Pont}, {Vidal-Madjar}, \& {Sing}}]{Lecavelier2008}
{Lecavelier Des Etangs}, A., {Pont}, F., {Vidal-Madjar}, A., \& {Sing}, D.
  2008, \aap, 481, L83

\bibitem[{{Lecavelier des Etangs} {et~al.}(2004){Lecavelier des Etangs},
  {Vidal-Madjar}, {McConnell}, \& {H{\'e}brard}}]{Lecavelier2004}
{Lecavelier des Etangs}, A., {Vidal-Madjar}, A., {McConnell}, J.~C., \&
  {H{\'e}brard}, G. 2004, \aap, 418, L1

\bibitem[{{Linsky} {et~al.}(2010){Linsky}, {Yang}, {France}, {Froning},
  {Green}, {Stocke}, \& {Osterman}}]{Linsky2010}
{Linsky}, J.~L., {Yang}, H., {France}, K., {et~al.} 2010, \apj, 717, 1291

\bibitem[{{Louden} \& {Wheatley}(2015)}]{Louden2015}
{Louden}, T. \& {Wheatley}, P.~J. 2015, \apjl, 814, L24

\bibitem[{{Lundkvist} {et~al.}(2016){Lundkvist}, {Kjeldsen}, {Albrecht},
  {Davies}, {Basu}, {Huber}, {Justesen}, {Karoff}, {Silva Aguirre}, {van
  Eylen}, {Vang}, {Arentoft}, {Barclay}, {Bedding}, {Campante}, {Chaplin},
  {Christensen-Dalsgaard}, {Elsworth}, {Gilliland}, {Handberg}, {Hekker},
  {Kawaler}, {Lund}, {Metcalfe}, {Miglio}, {Rowe}, {Stello}, {Tingley}, \&
  {White}}]{Lundkvist2016}
{Lundkvist}, M.~S., {Kjeldsen}, H., {Albrecht}, S., {et~al.} 2016, Nature
  Communications, 7, 11201

\bibitem[{{Mandel} \& {Agol}(2002)}]{Mandel2002}
{Mandel}, K. \& {Agol}, E. 2002, \apjl, 580, L171

\bibitem[{{Matsakos} {et~al.}(2015){Matsakos}, {Uribe}, \&
  {K{\"o}nigl}}]{Matsakos2015}
{Matsakos}, T., {Uribe}, A., \& {K{\"o}nigl}, A. 2015, \aap, 578, A6

\bibitem[{{Mauas} {et~al.}(2005){Mauas}, {Andretta}, {Falchi}, {Falciani},
  {Teriaca}, \& {Cauzzi}}]{Mauas2005}
{Mauas}, P.~J.~D., {Andretta}, V., {Falchi}, A., {et~al.} 2005, \apj, 619, 604

\bibitem[{{McCullough} {et~al.}(2014){McCullough}, {Crouzet}, {Deming}, \&
  {Madhusudhan}}]{McCullough2014}
{McCullough}, P.~R., {Crouzet}, N., {Deming}, D., \& {Madhusudhan}, N. 2014,
  \apj, 791, 55

\bibitem[{{Moutou} {et~al.}(2003){Moutou}, {Coustenis}, {Schneider}, {Queloz},
  \& {Mayor}}]{Moutou2003}
{Moutou}, C., {Coustenis}, A., {Schneider}, J., {Queloz}, D., \& {Mayor}, M.
  2003, \aap, 405, 341

\bibitem[{{Nortmann} {et~al.}(2018){Nortmann}, {Pall{\'e}}, {Salz},
  {Sanz-Forcada}, {Nagel}, {Alonso-Floriano}, {Czesla}, {Yan}, {Chen},
  {Snellen}, {Zechmeister}, {Schmitt}, {L{\'o}pez-Puertas}, {Casasayas-Barris},
  {Bauer}, {Amado}, {Caballero}, {Dreizler}, {Henning}, {Lamp{\'o}n}, {Montes},
  {Molaverdikhani}, {Quirrenbach}, {Reiners}, {Ribas}, {S{\'a}nchez-L{\'o}pez},
  {Schneider}, \& {Zapatero Osorio}}]{Lisa2018}
{Nortmann}, L., {Pall{\'e}}, E., {Salz}, M., {et~al.} 2018, Science,
  DOI:10.1126/science.aat5348

\bibitem[{{Ohta} {et~al.}(2005){Ohta}, {Taruya}, \& {Suto}}]{Ohta2005}
{Ohta}, Y., {Taruya}, A., \& {Suto}, Y. 2005, \apj, 622, 1118

\bibitem[{{Oklop{\v c}i{\'c}} \& {Hirata}(2018)}]{Oklopcic2018}
{Oklop{\v c}i{\'c}}, A. \& {Hirata}, C.~M. 2018, \apjl, 855, L11

\bibitem[{{Pillitteri} {et~al.}(2014){Pillitteri}, {Wolk}, {Lopez-Santiago},
  {G{\"u}nther}, {Sciortino}, {Cohen}, {Kashyap}, \& {Drake}}]{Pilliteri2014}
{Pillitteri}, I., {Wolk}, S.~J., {Lopez-Santiago}, J., {et~al.} 2014, \apj,
  785, 145

\bibitem[{{Pino} {et~al.}(2018){Pino}, {Ehrenreich}, {Wyttenbach}, {Bourrier},
  {Nascimbeni}, {Heng}, {Grimm}, {Lovis}, {Malik}, {Pepe}, \&
  {Piotto}}]{Pino2018}
{Pino}, L., {Ehrenreich}, D., {Wyttenbach}, A., {et~al.} 2018, \aap, 612, A53

\bibitem[{{Piskunov} \& {Valenti}(2002)}]{Piskunov2002}
{Piskunov}, N.~E. \& {Valenti}, J.~A. 2002, \aap, 385, 1095

\bibitem[{{Pont} {et~al.}(2008){Pont}, {Knutson}, {Gilliland}, {Moutou}, \&
  {Charbonneau}}]{Pont2008}
{Pont}, F., {Knutson}, H., {Gilliland}, R.~L., {Moutou}, C., \& {Charbonneau},
  D. 2008, \mnras, 385, 109

\bibitem[{{Pont} {et~al.}(2013){Pont}, {Sing}, {Gibson}, {Aigrain}, {Henry}, \&
  {Husnoo}}]{Pont2013}
{Pont}, F., {Sing}, D.~K., {Gibson}, N.~P., {et~al.} 2013, \mnras, 432, 2917

\bibitem[{{Poppenhaeger} {et~al.}(2013){Poppenhaeger}, {Schmitt}, \&
  {Wolk}}]{Poppenhaeger2013}
{Poppenhaeger}, K., {Schmitt}, J.~H.~M.~M., \& {Wolk}, S.~J. 2013, \apj, 773,
  62

\bibitem[{{Quirrenbach} {et~al.}(2018){Quirrenbach}, {Amado}, {Ribas},
  {Reiners}, {Caballero}, \& {et}}]{Quirrenbach2018}
{Quirrenbach}, A., {Amado}, P., {Ribas}, I., {et~al.} 2018, \procspie, 10702

\bibitem[{{Quirrenbach} {et~al.}(2016){Quirrenbach}, {Amado}, {Caballero},
  {Mundt}, {Reiners}, {Ribas}, {Seifert}, {Abril}, {Aceituno},
  {Alonso-Floriano}, {Anwand-Heerwart}, {Azzaro}, {Bauer}, {Barrado},
  {Becerril}, {Bejar}, {Benitez}, {Berdinas}, {Brinkm{\"o}ller}, {Cardenas},
  {Casal}, {Claret}, {Colom{\'e}}, {Cortes-Contreras}, {Czesla}, {Doellinger},
  {Dreizler}, {Feiz}, {Fernandez}, {Ferro}, {Fuhrmeister}, {Galadi},
  {Gallardo}, {G{\'a}lvez-Ortiz}, {Garcia-Piquer}, {Garrido}, {Gesa},
  {G{\'o}mez Galera}, {Gonz{\'a}lez Hern{\'a}ndez}, {Gonzalez Peinado},
  {Gr{\"o}zinger}, {Gu{\`a}rdia}, {Guenther}, {de Guindos}, {Hagen}, {Hatzes},
  {Hauschildt}, {Helmling}, {Henning}, {Hermann}, {Hern{\'a}ndez Arabi},
  {Hern{\'a}ndez Casta{\~n}o}, {Hern{\'a}ndez Hernando}, {Herrero}, {Huber},
  {Huber}, {Huke}, {Jeffers}, {de Juan}, {Kaminski}, {Kehr}, {Kim}, {Klein},
  {Kl{\"u}ter}, {K{\"u}rster}, {Lafarga}, {Lara}, {Lamert}, {Laun},
  {Launhardt}, {Lemke}, {Lenzen}, {Llamas}, {Lopez del Fresno},
  {L{\'o}pez-Puertas}, {L{\'o}pez-Santiago}, {Lopez Salas}, {Magan
  Madinabeitia}, {Mall}, {Mandel}, {Mancini}, {Marin Molina}, {Maroto
  Fern{\'a}ndez}, {Mart{\'{\i}}n}, {Mart{\'{\i}}n-Ruiz}, {Marvin}, {Mathar},
  {Mirabet}, {Montes}, {Morales}, {Morales Mu{\~n}oz}, {Nagel}, {Naranjo},
  {Nowak}, {Palle}, {Panduro}, {Passegger}, {Pavlov}, {Pedraz}, {Perez},
  {P{\'e}rez-Medialdea}, {Perger}, {Pluto}, {Ram{\'o}n}, {Rebolo}, {Redondo},
  {Reffert}, {Reinhart}, {Rhode}, {Rix}, {Rodler}, {Rodr{\'{\i}}guez},
  {Rodr{\'{\i}}guez L{\'o}pez}, {Rohloff}, {Rosich}, {Sanchez Carrasco},
  {Sanz-Forcada}, {Sarkis}, {Sarmiento}, {Sch{\"a}fer}, {Schiller}, {Schmidt},
  {Schmitt}, {Sch{\"o}fer}, {Schweitzer}, {Shulyak}, {Solano}, {Stahl},
  {Storz}, {Tabernero}, {Tala}, {Tal-Or}, {Ulbrich}, {Veredas}, {Vico Linares},
  {Vilardell}, {Wagner}, {Winkler}, {Zapatero Osorio}, {Zechmeister},
  {Ammler-von Eiff}, {Anglada-Escud{\'e}}, {del Burgo}, {Garcia-Vargas},
  {Klutsch}, {Lizon}, {Lopez-Morales}, {Ofir}, {P{\'e}rez-Calpena}, {Perryman},
  {S{\'a}nchez-Blanco}, {Strachan}, {St{\"u}rmer}, {Su{\'a}rez}, {Trifonov},
  {Tulloch}, \& {Xu}}]{Quirrenbach2016}
{Quirrenbach}, A., {Amado}, P.~J., {Caballero}, J.~A., {et~al.} 2016, in
  \procspie, Vol. 9908, Ground-based and Airborne Instrumentation for Astronomy
  VI, 990812

\bibitem[{{Redfield} {et~al.}(2008){Redfield}, {Endl}, {Cochran}, \&
  {Koesterke}}]{Redfield2008}
{Redfield}, S., {Endl}, M., {Cochran}, W.~D., \& {Koesterke}, L. 2008, \apjl,
  673, L87

\bibitem[{{Rodler} {et~al.}(2013){Rodler}, {K{\"u}rster}, \&
  {Barnes}}]{Rodler2013}
{Rodler}, F., {K{\"u}rster}, M., \& {Barnes}, J.~R. 2013, \mnras, 432, 1980

\bibitem[{{Salz} {et~al.}(2016){Salz}, {Czesla}, {Schneider}, \&
  {Schmitt}}]{Salz2016}
{Salz}, M., {Czesla}, S., {Schneider}, P.~C., \& {Schmitt}, J.~H.~M.~M. 2016,
  \aap, 586, A75

\bibitem[{{Sanz-Forcada} \& {Dupree}(2008)}]{Sanz2008}
{Sanz-Forcada}, J. \& {Dupree}, A.~K. 2008, \aap, 488, 715

\bibitem[{{Sanz-Forcada} {et~al.}(2011){Sanz-Forcada}, {Micela}, {Ribas},
  {Pollock}, {Eiroa}, {Velasco}, {Solano}, \&
  {Garc{\'{\i}}a-{\'A}lvarez}}]{Sanz2011}
{Sanz-Forcada}, J., {Micela}, G., {Ribas}, I., {et~al.} 2011, \aap, 532, A6

\bibitem[{{Schmitt} {et~al.}(1995){Schmitt}, {Fleming}, \&
  {Giampapa}}]{schmitt1995}
{Schmitt}, J.~H.~M.~M., {Fleming}, T.~A., \& {Giampapa}, M.~S. 1995, \apj, 450,
  392

\bibitem[{{Seager} \& {Sasselov}(2000)}]{Seager2000}
{Seager}, S. \& {Sasselov}, D.~D. 2000, \apj, 537, 916

\bibitem[{{Showman} {et~al.}(2013){Showman}, {Fortney}, {Lewis}, \&
  {Shabram}}]{Showman2013}
{Showman}, A.~P., {Fortney}, J.~J., {Lewis}, N.~K., \& {Shabram}, M. 2013,
  \apj, 762, 24

\bibitem[{{Sing} {et~al.}(2009){Sing}, {D{\'e}sert}, {Lecavelier Des Etangs},
  {Ballester}, {Vidal-Madjar}, {Parmentier}, {Hebrard}, \& {Henry}}]{Sing2009}
{Sing}, D.~K., {D{\'e}sert}, J.-M., {Lecavelier Des Etangs}, A., {et~al.} 2009,
  \aap, 505, 891

\bibitem[{{Sing} {et~al.}(2016){Sing}, {Fortney}, {Nikolov}, {Wakeford},
  {Kataria}, {Evans}, {Aigrain}, {Ballester}, {Burrows}, {Deming},
  {D{\'e}sert}, {Gibson}, {Henry}, {Huitson}, {Knutson}, {Lecavelier Des
  Etangs}, {Pont}, {Showman}, {Vidal-Madjar}, {Williamson}, \&
  {Wilson}}]{Sing2016}
{Sing}, D.~K., {Fortney}, J.~J., {Nikolov}, N., {et~al.} 2016, \nat, 529, 59

\bibitem[{{Sing} {et~al.}(2011){Sing}, {Pont}, {Aigrain}, {Charbonneau},
  {D{\'e}sert}, {Gibson}, {Gilliland}, {Hayek}, {Henry}, {Knutson}, {Lecavelier
  Des Etangs}, {Mazeh}, \& {Shporer}}]{Sing2011}
{Sing}, D.~K., {Pont}, F., {Aigrain}, S., {et~al.} 2011, \mnras, 416, 1443

\bibitem[{{Smette} {et~al.}(2015){Smette}, {Sana}, {Noll}, {Horst}, {Kausch},
  {Kimeswenger}, {Barden}, {Szyszka}, {Jones}, {Gallenne}, {Vinther},
  {Ballester}, \& {Taylor}}]{Smette2015}
{Smette}, A., {Sana}, H., {Noll}, S., {et~al.} 2015, \aap, 576, A77

\bibitem[{{Spake} {et~al.}(2018){Spake}, {Sing}, {Evans}, {Oklop{\v c}i{\'c}},
  {Bourrier}, {Kreidberg}, {Rackham}, {Irwin}, {Ehrenreich}, {Wyttenbach},
  {Wakeford}, {Zhou}, {Chubb}, {Nikolov}, {Goyal}, {Henry}, {Williamson},
  {Blumenthal}, {Anderson}, {Hellier}, {Charbonneau}, {Udry}, \&
  {Madhusudhan}}]{Spake2018}
{Spake}, J.~J., {Sing}, D.~K., {Evans}, T.~M., {et~al.} 2018, \nat, 557, 68

\bibitem[{{Triaud} {et~al.}(2009){Triaud}, {Queloz}, {Bouchy}, {Moutou},
  {Collier Cameron}, {Claret}, {Barge}, {Benz}, {Deleuil}, {Guillot},
  {H{\'e}brard}, {Lecavelier Des {\'E}tangs}, {Lovis}, {Mayor}, {Pepe}, \&
  {Udry}}]{Triaud2009}
{Triaud}, A.~H.~M.~J., {Queloz}, D., {Bouchy}, F., {et~al.} 2009, \aap, 506,
  377

\bibitem[{{Vidal-Madjar} {et~al.}(2004){Vidal-Madjar}, {D{\'e}sert},
  {Lecavelier des Etangs}, {H{\'e}brard}, {Ballester}, {Ehrenreich}, {Ferlet},
  {McConnell}, {Mayor}, \& {Parkinson}}]{Vidal2004}
{Vidal-Madjar}, A., {D{\'e}sert}, J.-M., {Lecavelier des Etangs}, A., {et~al.}
  2004, \apjl, 604, L69

\bibitem[{{Vidal-Madjar} {et~al.}(2013){Vidal-Madjar}, {Huitson}, {Bourrier},
  {D{\'e}sert}, {Ballester}, {Lecavelier des Etangs}, {Sing}, {Ehrenreich},
  {Ferlet}, {H{\'e}brard}, \& {McConnell}}]{Vidal2013}
{Vidal-Madjar}, A., {Huitson}, C.~M., {Bourrier}, V., {et~al.} 2013, \aap, 560,
  A54

\bibitem[{{Vidal-Madjar} {et~al.}(2003){Vidal-Madjar}, {Lecavelier des Etangs},
  {D{\'e}sert}, {Ballester}, {Ferlet}, {H{\'e}brard}, \& {Mayor}}]{Vidal2003}
{Vidal-Madjar}, A., {Lecavelier des Etangs}, A., {D{\'e}sert}, J.-M., {et~al.}
  2003, \nat, 422, 143

\bibitem[{{Vidal-Madjar} {et~al.}(2008){Vidal-Madjar}, {Lecavelier des Etangs},
  {D{\'e}sert}, {Ballester}, {Ferlet}, {H{\'e}brard}, \& {Mayor}}]{Vidal2008}
{Vidal-Madjar}, A., {Lecavelier des Etangs}, A., {D{\'e}sert}, J.-M., {et~al.}
  2008, \apjl, 676, L57

\bibitem[{{Watson} {et~al.}(1981){Watson}, {Donahue}, \& {Walker}}]{Watson1981}
{Watson}, A.~J., {Donahue}, T.~M., \& {Walker}, J.~C.~G. 1981, \icarus, 48, 150

\bibitem[{{Wyttenbach} {et~al.}(2015){Wyttenbach}, {Ehrenreich}, {Lovis},
  {Udry}, \& {Pepe}}]{Wyttenbach2015}
{Wyttenbach}, A., {Ehrenreich}, D., {Lovis}, C., {Udry}, S., \& {Pepe}, F.
  2015, \aap, 577, A62

\bibitem[{{Yan} \& {Henning}(2018)}]{Yan2018}
{Yan}, F. \& {Henning}, T. 2018, Nature Astronomy, 2, 714

\bibitem[{{Zarro} \& {Zirin}(1986)}]{Zarro1986}
{Zarro}, D.~M. \& {Zirin}, H. 1986, \apj, 304, 365

\bibitem[{{Zechmeister} {et~al.}(2018){Zechmeister}, {Reiners}, {Amado},
  {Azzaro}, {Bauer}, {B{\'e}jar}, {Caballero}, {Guenther}, {Hagen}, {Jeffers},
  {Kaminski}, {K{\"u}rster}, {Launhardt}, {Montes}, {Morales}, {Quirrenbach},
  {Reffert}, {Ribas}, {Seifert}, {Tal-Or}, \& {Wolthoff}}]{Zechmeister2018}
{Zechmeister}, M., {Reiners}, A., {Amado}, P.~J., {et~al.} 2018, \aap, 609, A12

\end{thebibliography}

\appendix

%_______________________________________________________________________________
%_______________________________________________________________________________
\section{Complementary figures}\label{Sect:APP}
%_______________________________________________________________________________
%_______________________________________________________________________________

\begin{figure*}[t]
  \centering
  \includegraphics[draft=false, width=\hsize, trim = 2.5mm 2.5mm 3mm 0mm, clip]{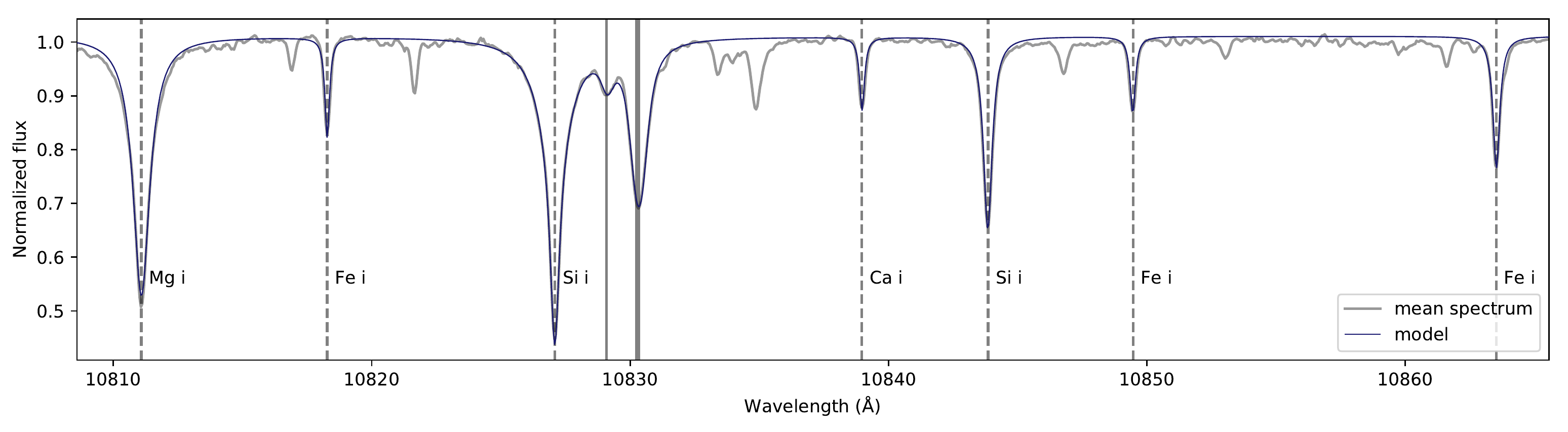}
  \caption{Spectral region of the \lmain{} lines. The gray line is the average stellar
           spectrum, the dark-blue line is a model that includes the seven strongest  stellar lines in the direct vicinity of the triplet lines used for the correction of the stellar rest frame. The position of the \hei{} triplet is marked by vertical solid lines; the other seven spectral lines are labeled and marked by dashed lines.
           }
  \label{fig:star_rf}
\end{figure*}

\begin{figure}[h]
  \centering
  \includegraphics[draft=false, width=\hsize, trim = 2.5mm 13mm 3mm 0mm, clip]{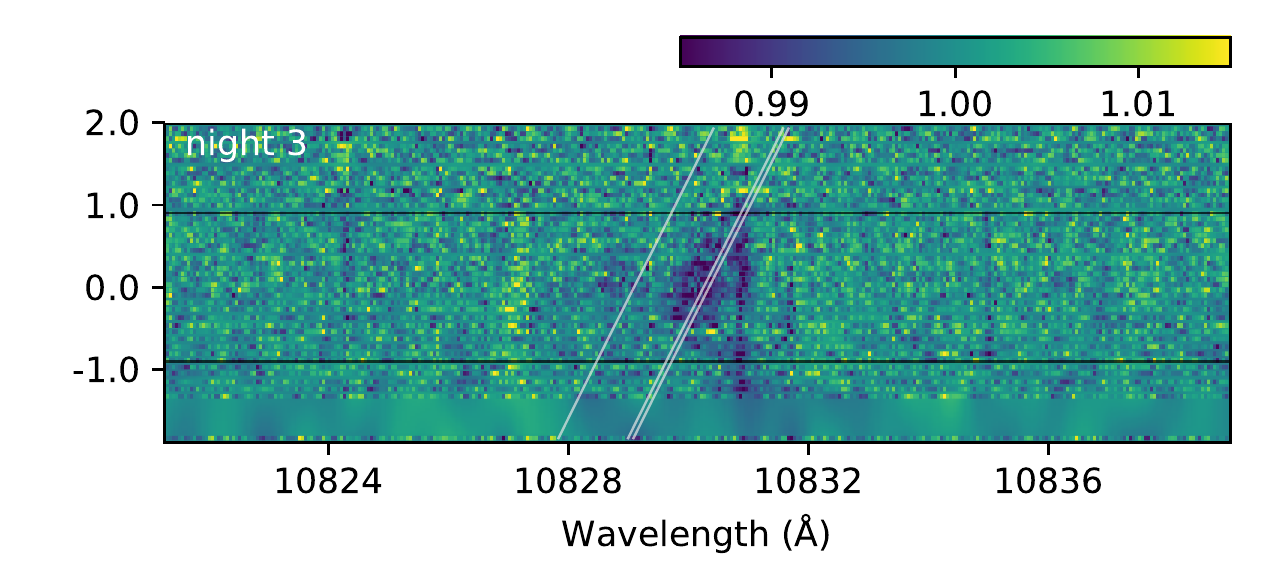}\\
  \includegraphics[draft=false, width=\hsize, trim = 2.5mm 13mm 3mm 14mm, clip]{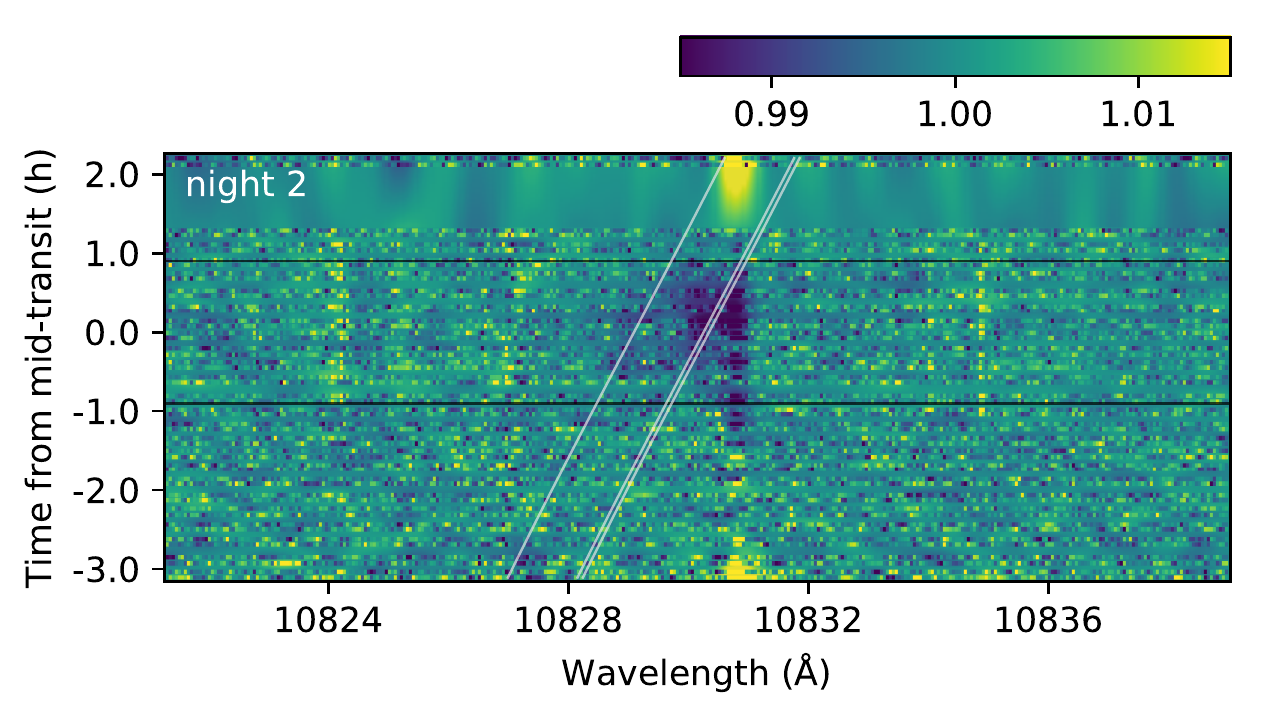}\\
  \includegraphics[draft=false, width=\hsize, trim = 2.5mm 0mm 3mm 12.5mm, clip]{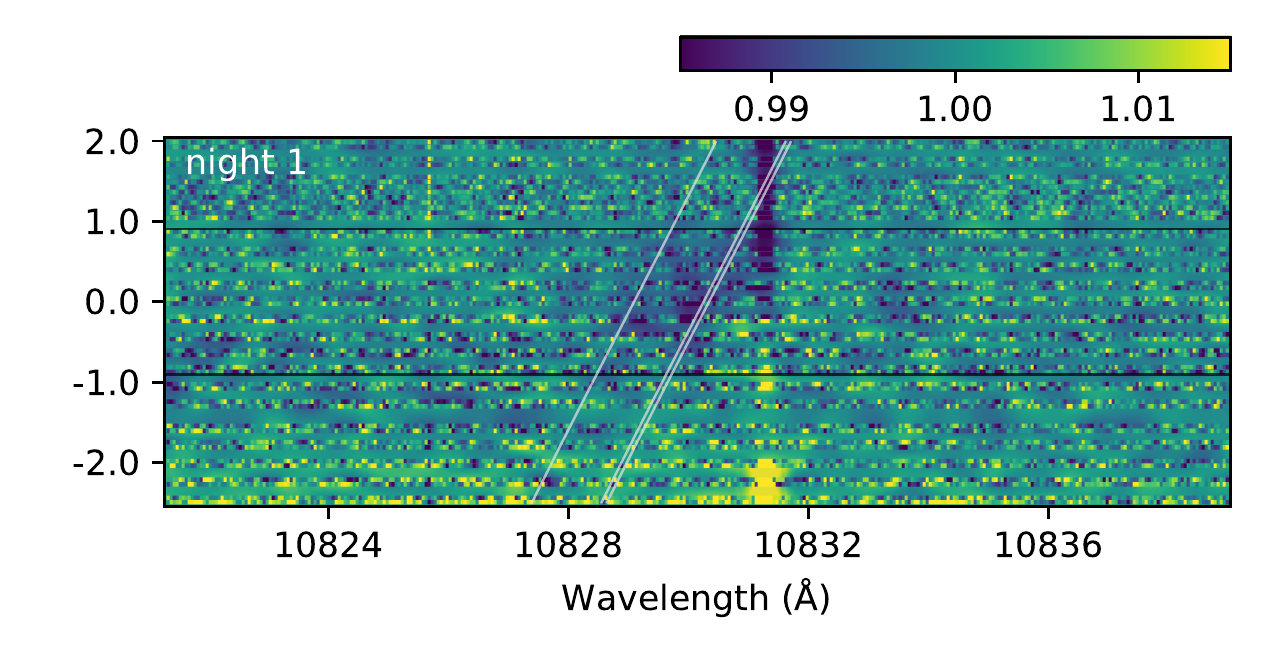}
  \caption{Residual spectra in the stellar rest frame before correction of
           the telluric emission lines (similar to Fig.~\ref{fig:tsp}).
           }
  \label{fig:tsp_ntc}
\end{figure}

\begin{figure}[h]
  \centering
  \includegraphics[width=0.815\hsize, trim = 2.5mm 2.5mm 3mm 13mm, clip]{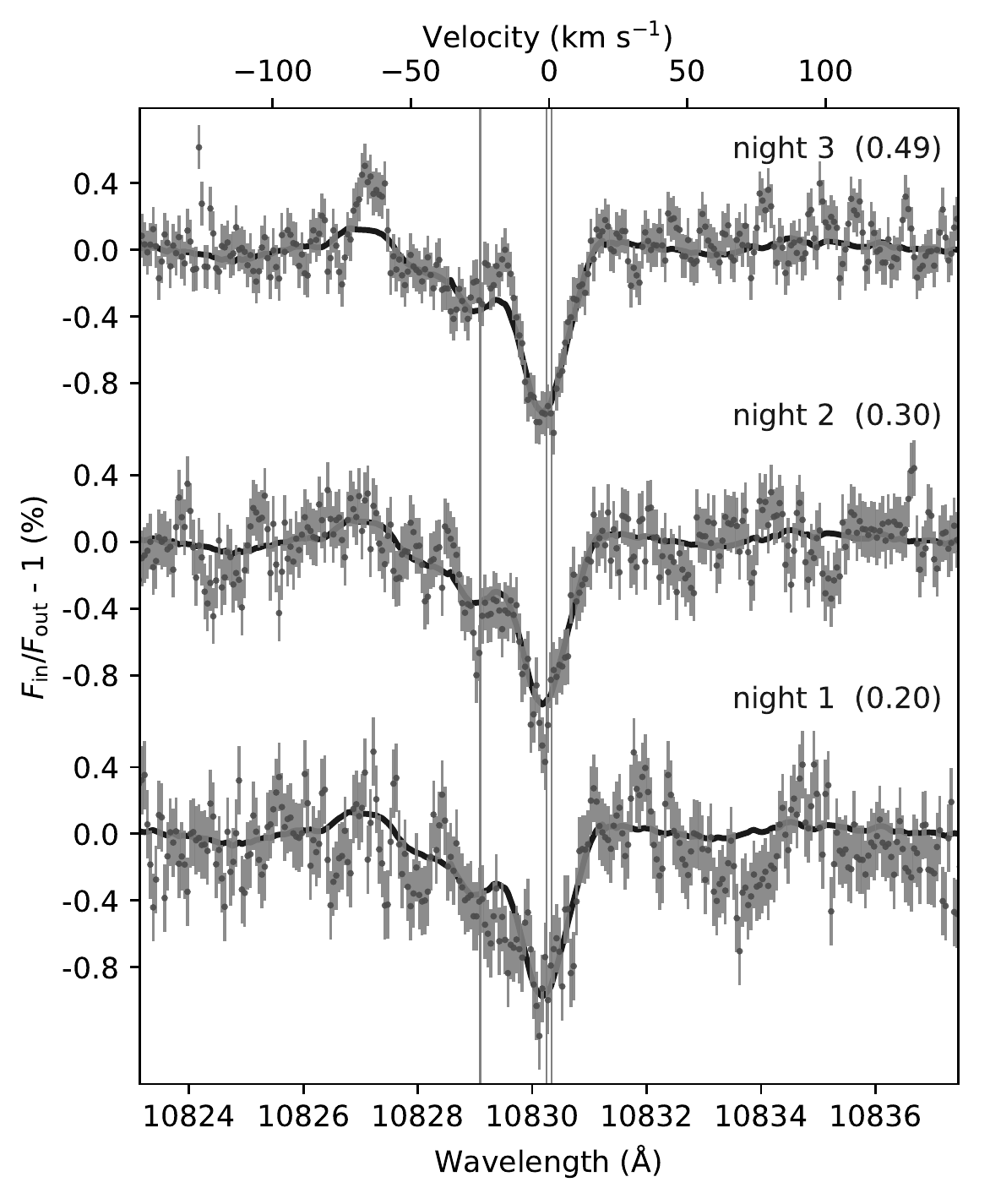}\\\vspace{5pt}
  \includegraphics[width=0.815\hsize, trim =2.5mm 2.5mm 3mm 12.2mm, clip]{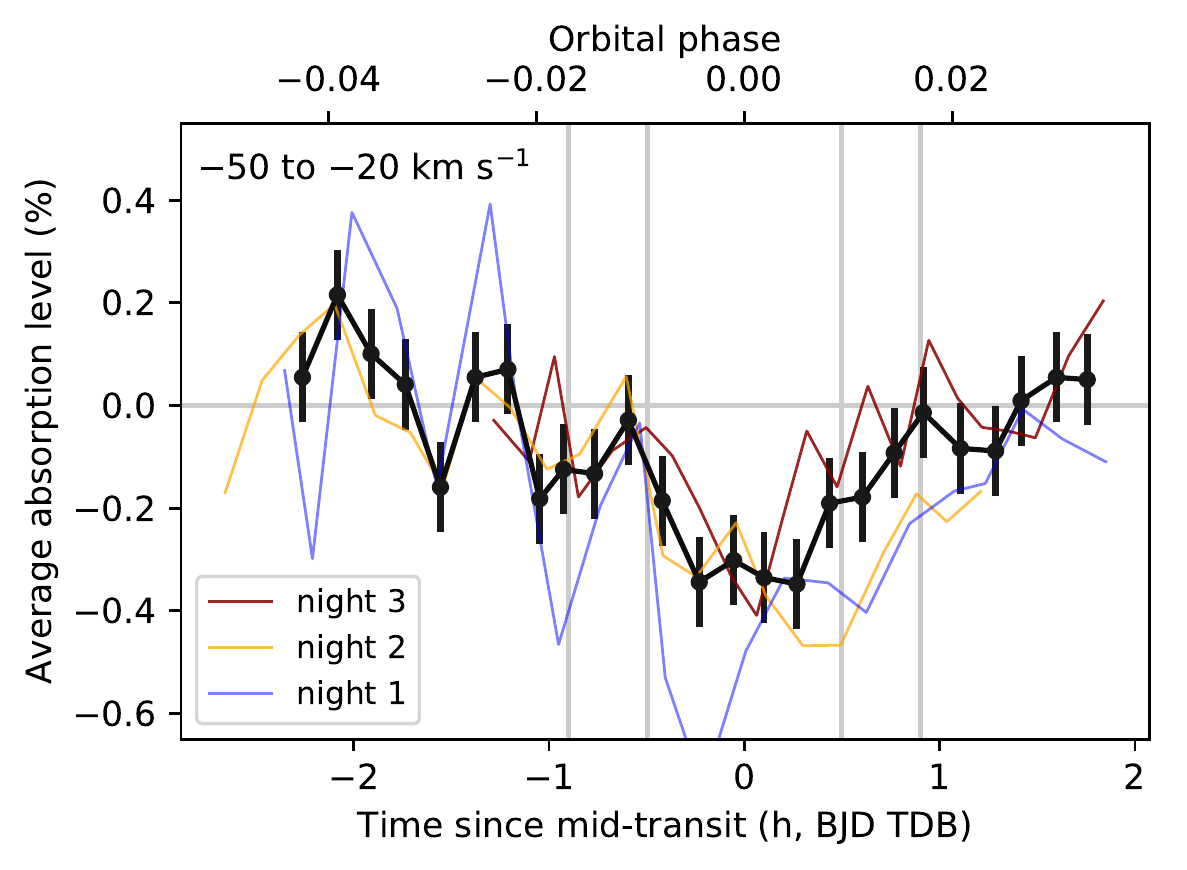}
  \caption{{\it Top panel:} Transmission spectra of the three individual
           transits in the planetary rest frame.
           The \hei{} triplet is indicated by vertical lines. 
           In parentheses we provide the weights for computing the weighted
           mean absorption profile.
           For comparison, a smoothed version of the mean absorption profile is given
           by the black solid line.
           For a discussion see Sects.~\ref{SectAnalysis} and \ref{Sect:APP_absdepth}.
           {\it Bottom panel:} 
           Same as Fig.~\ref{fig:lc}, but for the \lminor{} line.
           }
  \label{fig:var+lc2}
\end{figure}

Figure~\ref{fig:star_rf} shows the stellar lines used to correct the alignment of all spectra in the stellar rest frame. Figure~\ref{fig:tsp_ntc} shows the residual spectra as depicted in Fig.~\ref{fig:tsp}, but without the removal of telluric emission lines. In the top panel of Fig.~\ref{fig:var+lc2}, we display the average residual spectra during the mid-transit phase of the three individual transit observations. Finally, the bottom panel of Fig.~\ref{fig:var+lc2} depicts light curves centered on the minor component.

\newpage\mbox{} \newpage
\section{Calcium IRT lines}\label{Sect:APP_IRT}

\begin{figure}[h]
  \centering
  \includegraphics[width=\hsize, trim = 2.5mm 2.5mm 3mm 2mm, clip]{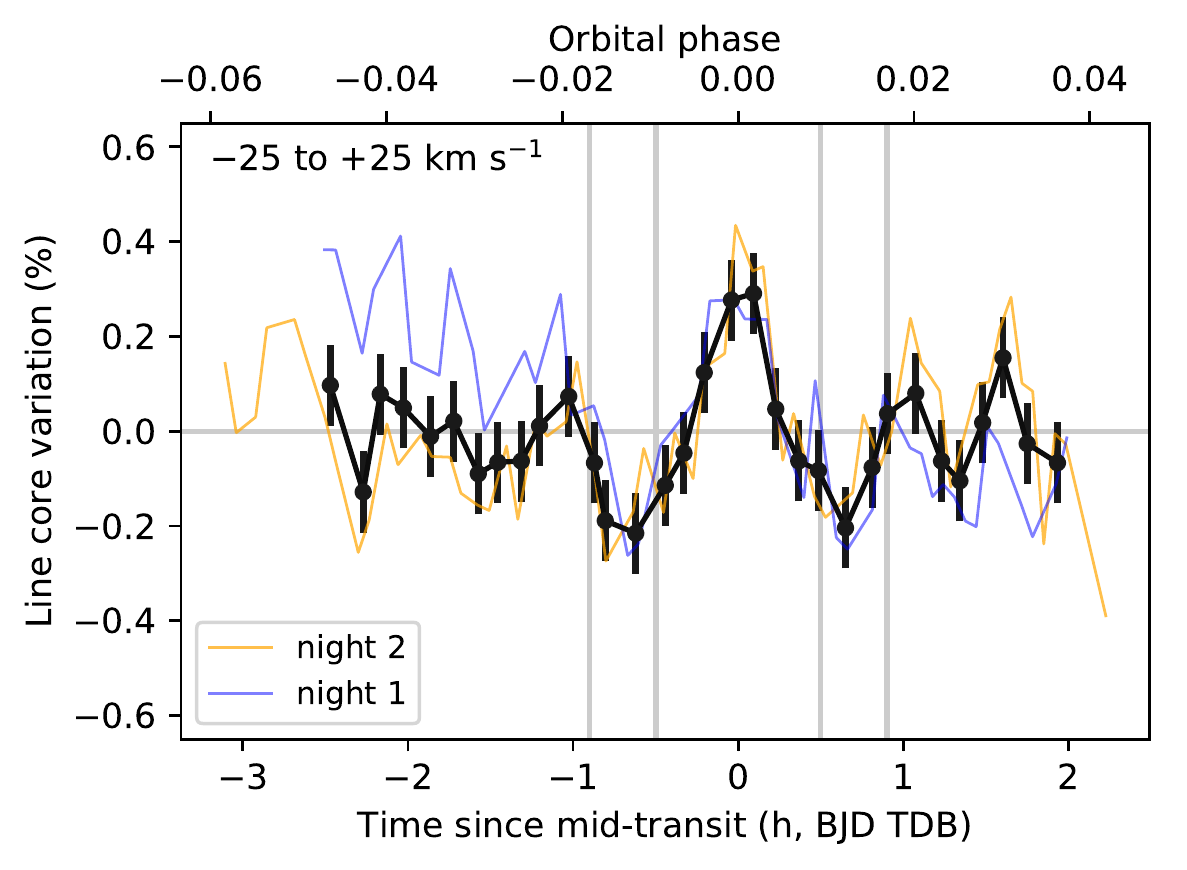}
  \caption{Light curves of the averaged three Ca~IRT lines during the first two
           transit nights.
           The bold curve shows the mean light curve after correction of
           activity trends.
           Vertical lines mark the contact points of the optical transit.
           }
  \label{fig:cairt}
\end{figure}

To monitor the activity evolution of the host star, we checked the time evolution of the three \ion{Ca}{ii} IRT lines. We averaged the residual flux in a $\pm$\,25~\kms{} range in the line cores and then averaged the three individual lines. The resulting light curves are shown in Fig.~\ref{fig:cairt} for  nights~1 and 2. The visual channel failed on night~3.
Night~2 shows a slight activity trend that was fit linearly and subtracted before computing the mean IRT transit light curve. The light curve has the usual shape expected from center-to-limb variation in the stellar lines \citep{Czesla2015}. We find no indications of flaring activity, and in particular, we do not find evidence that the planet crossed unusually strong active regions during the two transit nights,
which would cause larger excursions from the average light curve. This is consistent with our \hei{} analysis.

%_______________________________________________________________________________
%_______________________________________________________________________________
\section{Rossiter--McLaughlin effect}\label{Sect:RM}
\begin{figure}[h]
  \centering
  \includegraphics[width=\hsize, trim = 2.5mm 2.5mm 3mm 2mm, clip]{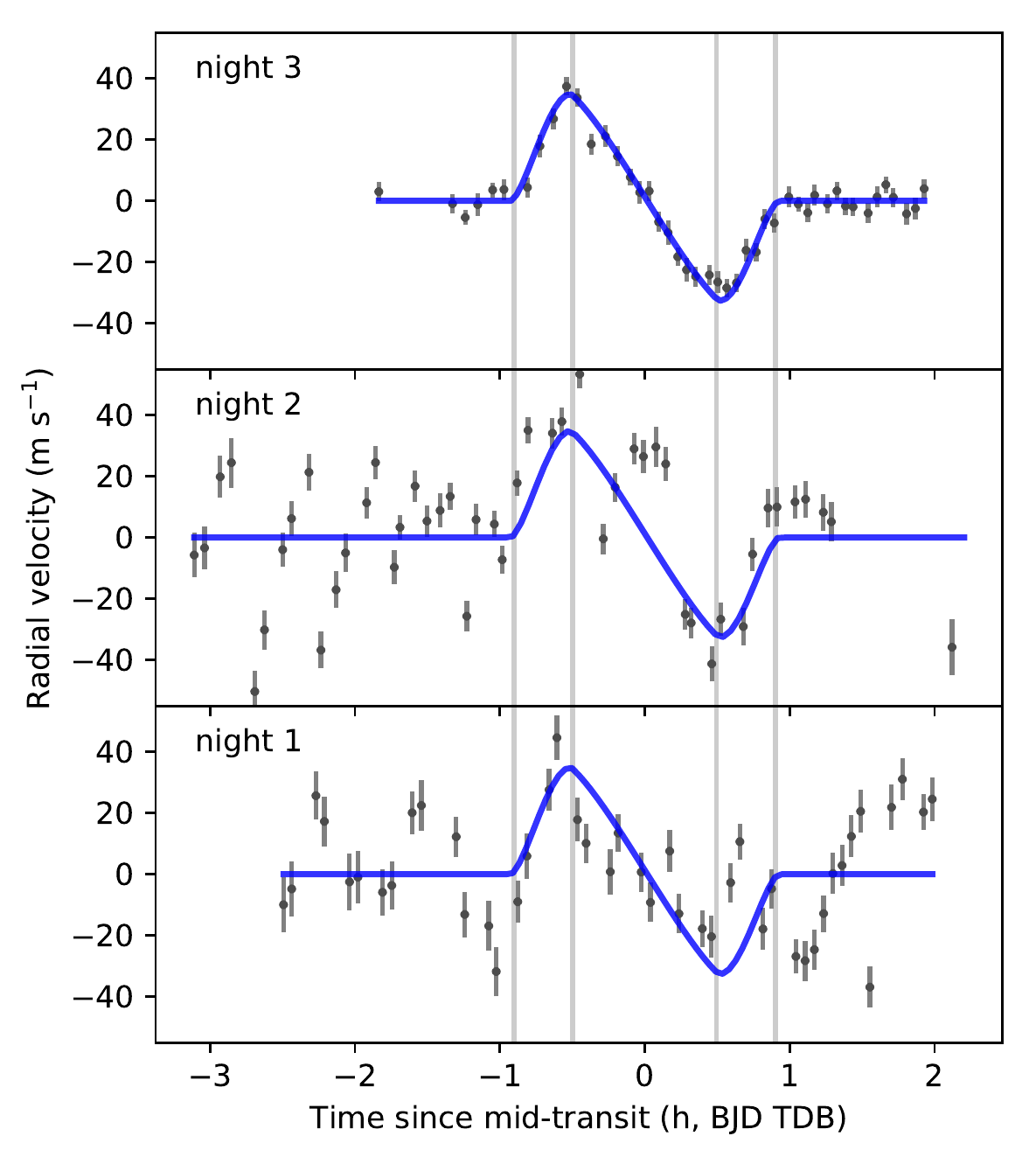}
  \caption{Radial velocity measurements of the NIR channel.
           A model of the RME is shown by the blue line.
           The four contact points are marked by vertical lines.
           }
  \label{fig:shifts}
\end{figure}

%_______________________________________________________________________________
%_______________________________________________________________________________
We derived relative radial velocity measurements for the NIR channel of CARMENES using SERVAL \citep{Zechmeister2018}. As described in Sect.~\ref{tab:cont_norm}, the instrument can exhibit nightly drifts with the  configuration used. Therefore, we corrected linear trends from the radial velocity measurements. The resulting RVs are shown in Fig.~\ref{fig:shifts}. The first two nights exhibit a large scatter. These observations were taken shortly after the instrument commissioning and the NIR spectrograph appears not to be fully stabilized yet \citep{Quirrenbach2018}.

We used only night~3 to fit the Rossiter--McLaughlin effect (RME) following the prescription of \citet{Ohta2005}. The posterior distributions were explored with an MCMC chain of $10^5$ steps with a burn-in of $10^4$ steps. The reference mid-transit time, period, and semimajor axis were set-up with Gaussian priors according to the values in Table.~\ref{TabPara}. These parameters were not further confined by our data. The linear limb-darkening parameter was fixed to 0.43, which is the value calculated for the $J$ band and a \hdo{}-like star by \citet{Claret2011}. The inclination of the stellar rotation axis was fixed to 92\degr{} \citep{Cegla2016}.

The remaining free parameters were the radius ratio, the rotation velocity of the host $\Omega$, the inclination of the orbit $i$, and the obliquity $\lambda$. For the obliquity and the inclination, we find $\lambda=-1.6 \pm 1.1$\degr{} and $i=85.77 \pm 0.11$\degr{}, consistent with literature values \citep[e.g.,][]{Triaud2009, Cegla2016, CasasayasBarris2017}. 
With a value of $0.162\pm0.024$, the radius ratio is only weakly confined by our data and consistent with all the literature values. For the angular rotation we find $\Omega = 0.40 \pm 0.11$~rad/day, which converts to a rotation period of $15.8\pm4.6$~d. This is somewhat larger than but still consistent with the well-determined value of $11.953\pm0.009$ \citep{Henry2008}.

The slight discrepancy of the stellar rotation rate could be caused by differential rotation as noted by \citet{Cegla2016}. Also, a wavelength dependence of the radius ratio \citep{DiGloria2015} would affect the measured stellar rotation rate. We further note that our RVs were determined with a cross-correlation technique, while the model assumed a line centroid method. A detailed analysis of the chromatic RME including the visual channel of CARMENES is beyond the scope of this paper.

Neither the stellar rotation period nor the radius ratio have any impact on our main conclusions here. The RME confirms the  ephemeris used. The large scatter of the RVs during the first two observation nights shows that the instrument was less stable, and supports our analysis procedure to put the greatest weight on the observations during night~3.

\end{document}